\documentclass[11pt]{article}

\usepackage{fullpage}

\usepackage[utf8]{inputenc}
\usepackage{amsmath}
\usepackage{amsthm}
\usepackage{bbm}
\usepackage{amsfonts}
\usepackage{amssymb}
\usepackage{units}
\usepackage{xcolor}
\usepackage{graphicx}
\usepackage[ruled,linesnumbered]{algorithm2e}
\usepackage{comment}
\usepackage{xfrac}
\usepackage{tikz}
\usetikzlibrary{arrows.meta, positioning}
\usepackage{subcaption}
\let\oldnl\nl
\newcommand{\nonl}{\renewcommand{\nl}{\let\nl\oldnl}}

\usepackage{tikz}
\usetikzlibrary{arrows.meta, positioning}
\usepackage{enumitem}
\tikzset{directed/.style={->, >=Latex, thick}}

\newtheorem{theorem}{Theorem}[section]
\newtheorem{claim}{Claim}[theorem]
\newtheorem{lemma}[theorem]{Lemma}

\newtheorem{corollary}[theorem]{Corollary}
\newtheorem{definition}[theorem]{Definition}

\newcommand{\dist}{\mathrm{dist}}
\newcommand{\lra}{\leftrightarrows}
\newcommand{\D}{\mathcal{D}}
\newcommand{\eps}{\varepsilon}
\newcommand{\poly}{\operatorname{poly}}

\newcommand{\e}{\mathbb{E}}

\newcommand{\Dom}{\operatorname{dom}}

\title{Roundtrip All-Pairs Shortest Paths in Directed Graphs}

\author{
Shiri Chechik
\and
Gur Lifshitz
\thanks{Tel Aviv University, Israel.
E-mail: {\tt shiri.chechik@gmail.com, \tt gurlifshitz@gmail.com}.
}
}

\begin{document}

\maketitle

\begin{abstract}
 We study approximation algorithms for All-Pairs Shortest Paths (APSP) in directed graphs under the roundtrip metric. In undirected graphs, approximate APSP has been studied extensively, yielding a rich landscape of tradeoffs between approximation stretch and running time, from the classical \(O(mn)\) exact algorithm toward the \(\Omega(n^2)\) output-size barrier. In directed graphs, however, even finite-factor approximation of ordinary one-way distances must distinguish reachable from unreachable vertex pairs, and therefore computes transitive closure, a problem equivalent to Boolean matrix multiplication (BMM) under standard reductions.
The roundtrip distance, defined as $\text{dist}_G(u \leftrightarrows v) = \text{dist}_G(u \to v) + \text{dist}_G(v \to u)$, offers a natural and robust relaxation. It measures the cost of traveling from $u$ to $v$ and back, satisfies the triangle inequality, and is inherently symmetric. While this metric has become central to the study of directed roundtrip spanners and compact routing, the fundamental problem of all-pairs roundtrip distance approximation has remained largely unexplored.
In this work, we present a comprehensive study of roundtrip APSP and distance
oracles, presenting several new lower and upper bounds. On the algorithmic side,
our main result is a combinatorial algorithm that computes a stretch-$3$
approximation to all-pairs roundtrip distances in $\widetilde{O}(n^2 + \min\{m\sqrt{n},\, m^{1/3}n^{5/3}\})$ time. For edge weights in $[1,W]$, we also obtain a $(3+\varepsilon)$-approximation
in $\widetilde O((n^{2.037}+nm^{2/3})\log(nW)/\varepsilon)$ time
and a $(4+\varepsilon)$-approximation in
$\widetilde O(n^{2.032}\log(nW)/\varepsilon)$ time using fast matrix
multiplication.
On the hardness side, we show that any stretch below $3$ would imply
a breakthrough for BMM. We also obtain additional tradeoffs between
stretch and running time.


\end{abstract}

\section{Introduction}
The \emph{All-Pairs Shortest Paths} (APSP) problem is one of the most fundamental problems in graph algorithms. Given a graph $G=(V,E,w)$ with nonnegative edge weights, the goal is to compute the shortest-path distance between every ordered pair of vertices. For dense graphs with real nonnegative edge weights, the classical algorithm of Floyd--Warshall~\cite{cormen01introduction} runs in $O(n^3)$ time, and the fastest known exact algorithms improve over this bound only by $n^{o(1)}$ factors (see~\cite{Williams18}). A central hypothesis in fine-grained complexity asserts that no $O(n^{3-\varepsilon})$-time algorithm can solve APSP in edge-weighted graphs, for any fixed $\varepsilon>0$ (see~\cite{Virginia18}). Consequently, most progress toward faster algorithms has relied on either approximation techniques, or algebraic methods such as fast matrix multiplication combined with restrictions on the edge weights.

\paragraph{Approximating Undirected APSP}

The study of approximate APSP in undirected graphs dates back to the late
1990s. In \textbf{unweighted} graphs, Aingworth et al.~\cite{plusTwo} gave a
combinatorial \(+2\)-approximation in \(\widetilde O(n^{5/2})\) time, and
Dor, Halperin, and Zwick~\cite{Dor00} improved the running time to
\(\widetilde O(\min\{n^{3/2}m^{1/2},n^{7/3}\})\)\footnote{The $\widetilde O(\cdot)$ notation hides polylogarithmic factors in $n$.}. Roditty~\cite{Rod23} gave a combinatorial stretch-\(2\) algorithm in \(\widetilde O(n^{9/4})\) time, and Dory et
al.~\cite{DFK24} and independently Saha and Ye~\cite{SY24} obtained an
algebraic stretch-\(2\) algorithm in \(O(n^{2.0329})\) time.
For \textbf{weighted} graphs, Cohen and Zwick~\cite{CZ01} established the classical
tradeoffs, including stretch \(3\) in near-quadratic time, stretch \(7/3\) in
\(\widetilde O(n^{7/3})\) time, and stretch \(2\) in \(\widetilde O(n^{3/2}m^{1/2})\) time. Baswana and Kavitha~\cite{Baswana10} obtained a combinatorial
stretch-\(2\) algorithm running in \(\widetilde O(n^2+m\sqrt n)\) time. Further combinatorial and algebraic tradeoffs for weighted and unweighted graphs can be found in, e.g., \cite{Kavitha12,AR21,Gupta25,Jin26}.

\paragraph{Approximating Directed APSP}

The directed case is substantially less understood. For exact APSP with small integer weights, Alon, Galil, and Margalit~\cite{AGM97} gave an $O(n^{(\omega+3)/2})$-time algorithm, and Zwick~\cite{Zwick2002} later improved this to $O(n^{2.575})$ using rectangular matrix multiplication; this remains the fastest known exact bound even for unweighted directed graphs. Zwick also showed that, for polynomially bounded weights, a $(1+\varepsilon)$-approximation can be computed in $\widetilde O(n^\omega/\varepsilon)$ time. Roditty and Shapira~\cite{RS11} later obtained algebraic additive-error tradeoffs, achieving additive error $d_G(u,v)^p$ in time
\(
O\!\left(n^{2.575-\frac{p}{7.4-2.3p}}\right)
\) for every $0\le p\le 1$.

Any multiplicative approximation to directed distances must distinguish finite from infinite distances, and therefore computes transitive closure. Since transitive closure and Boolean matrix multiplication are equivalent under standard reductions~\cite{Fischer1971BooleanMM,VW10}, such an approximation algorithm would also yield an algorithm for Boolean matrix multiplication with essentially the same running time. This motivates studying the roundtrip metric.


\paragraph{Roundtrip Distances}

A central concept in the analysis of directed graphs is the \emph{roundtrip distance}, introduced by Cowen and Wagner~\cite{CowenWagner00}. This metric is motivated by the observation that two vertices are often meaningfully close only if one can travel from one to the other and back at small total cost. For vertices \(u,v\in V(G)\), the roundtrip distance is defined as
\[
\dist_G(u \leftrightarrows v)
 =
 \dist_G(u\!\to\!v) + \dist_G(v\!\to\!u),
\]
that is, the total length of the shortest directed closed walk passing through both \(u\) and \(v\). This measure is symmetric and satisfies the triangle inequality, inducing a metric on every strongly connected component of \(G\).

One way to view this metric is through the complete undirected graph \(G^{\rm rt}\) on the vertex set \(V\), where the edge \(\{u,v\}\) has weight \(\dist_G(u \leftrightarrows v)\). The difficulty is that \(G^{\rm rt}\) is dense and implicit: computing its edge weights exactly is essentially the directed APSP problem. Thus, the challenge is to exploit the undirected metric structure of roundtrip distances without explicitly constructing the metric.

The roundtrip metric has played a central role in directed spanners and routing. Roditty, Thorup, and Zwick~\cite{RTZ08} initiated the study of roundtrip spanners for directed graphs, showing general \((2k+\varepsilon)\)-roundtrip spanners with \(\widetilde O_\varepsilon(kn^{1+1/k})\) edges, extending these constructions to compact roundtrip routing schemes, and proving that every weighted directed graph admits an optimal-size \(3\)-roundtrip spanner with \(O(n^{3/2})\) edges. Later work developed faster constructions of roundtrip spanners and closely related algorithms for directed-girth approximation.


Chechik, Liu, Rotem, and Sidford~\cite{roundtrip_spanners19} were the first to obtain a constant-factor approximation for the directed girth in subquadratic time, achieving a $3$-approximation in $\widetilde{O}(m\sqrt{n})$ time. Using closely related techniques, they also showed how to construct an $8$-roundtrip spanner with $\widetilde{O}(n^{3/2})$ edges in the same running time.

Dalirrooyfard and Vassilevska Williams~\cite{dalirrooyfard20} subsequently obtained a $(2+\varepsilon)$-approximation for weighted directed girth in $\widetilde{O}(m\sqrt{n})$ time, together with a $(5+\varepsilon)$-roundtrip spanner with $\widetilde{O}(n^{3/2})$ edges in the same running time.

Building on this line of work, Chechik and Lifshitz~\cite{CL21} gave a $2$-approximation for the girth in $\widetilde{O}(m\sqrt{n})$ time. They also presented a different algorithm that computes both a $2$-approximation for the girth and a $4$-roundtrip spanner with $\widetilde{O}(n^{3/2})$ edges in $\widetilde{O}(n^2)$ time.
More recently, Harbuzova, Jin, Vassilevska Williams, and Xu~\cite{HarbuzovaJinVWXu24} further extended these ideas to obtain a $3$-roundtrip spanner with $O(n^{3/2})$ edges in $\widetilde{O}(m\sqrt{n})$ time. The roundtrip metric has also found applications in compact routing and more roundtrip spanner constructions; see, e.g.,~\cite{RTZ08,
roundtrip_spanners18, roundtrip_spanners20b}.



\paragraph{Lower Bounds.}
For directed girth, Dalirrooyfard and Vassilevska Williams~\cite{dalirrooyfard20} showed that, under the $k$-Cycle Hypothesis, no $O(m^{2-\varepsilon})$-time algorithm can achieve a $(2-\delta)$-approximation, for any constants $\varepsilon,\delta > 0$.

\subsection{Our results}

We provide several algorithms for approximating all-pairs roundtrip distances in directed graphs, together with conditional barriers for stretch-$3$ approximation. Our algorithmic results are summarized in Table~\ref{tab:rt-apsp-summary}. At a high level, our upper bounds are based on two algorithmic ingredients. The first is a pivot-based distance-oracle construction using truncated outgoing searches. The second is a graph sparsification procedure that reduces the number of edges for faster distance computations. Each ingredient already gives a nontrivial roundtrip APSP approximation, and their combination yields our main combinatorial result: a $3$-APSP algorithm running in $\widetilde O(m^{1/3}n^{5/3}+n^2)$ time.

\paragraph{First ingredient: pivots and truncated searches.}

Our first ingredient is a stretch-$3$ roundtrip distance oracle. Querying the oracle in both directions then yields a stretch-$3$ approximation to the roundtrip distance.

\begin{theorem}
\label{th:3do}
There exists a $3$-roundtrip distance oracle of expected size $O(n^{3/2})$, constructible in expected time $\widetilde O(m\sqrt n)$ and with query time $O(\log n)$.
Specifically, for every pair of vertices $v,u \in V$, the oracle returns a value $\dist_{\mathcal{D}}(v,u)$ satisfying
\[
d_G(v \to u)
\le
\dist_{\mathcal{D}}(v,u)
\le
2d_G(v \to u) + d_G(u \to v).
\]
In particular, combining the estimates for $(v,u)$ and $(u,v)$ yields a $3$-approximation to the roundtrip distance $d_G(v\lra u)$.
\end{theorem}

This result can be viewed as a fast-construction analogue of the stretch-$3$ roundtrip spanner of Roditty, Thorup, and Zwick~\cite{RTZ08}. By querying this oracle on all vertex pairs, we obtain the following all-pairs approximation.

\begin{corollary}
\label{cor:3-apsp}
For any directed graph with nonnegative edge weights, one can compute a $3$-roundtrip APSP approximation in expected time
\(
\widetilde O(m\sqrt n+n^2).
\)
\end{corollary}

\paragraph{Second ingredient: sparsification.}

Our second ingredient is a fast sparsification procedure that preserves roundtrip distances approximately. After the sparsification procedure we run the distance oracle on a sparser graph and obtain a faster approximation for dense graphs.

\begin{theorem}
\label{th:4-apsp}
There exists a randomized combinatorial algorithm that computes a $4$-roundtrip APSP approximation in expected time
\(
\widetilde O(n^{9/4}).
\)
\end{theorem}

The main combinatorial upper bound is obtained by combining the two ingredients. Instead of running the sparsification step and then exploring the sparsified graph with the pivot machinery, we do both in parallel. This improves the stretch back to $3$ and beats the $\widetilde O(m\sqrt n+n^2)$ bound on dense graphs.

\begin{theorem}
\label{th:fast-3-apsp}
There exists a randomized combinatorial algorithm that, in expected time
\(
\widetilde O(m^{1/3}n^{5/3} + n^2),
\)
computes estimates $\dist_{\mathcal D}(v,u)$ for all ordered pairs
$v,u\in V$ satisfying
\[
d_G(v\to u)
\le
\dist_{\mathcal D}(v,u)
\le
2d_G(v\to u)+d_G(u\to v).
\]
In particular, this yields a $3$-approximation to all-pairs roundtrip distances.
\end{theorem}

For dense graphs, Theorem~\ref{th:fast-3-apsp} gives a $3$-APSP algorithm running in $\widetilde O(n^{7/3})$ time.

\paragraph{Lower bounds: reductions from Boolean matrix multiplication.}

We complement our algorithms with two reductions from Boolean Matrix
Multiplication. The first shows that improving the roundtrip stretch below
\(3\) would yield a corresponding improvement for BMM.

\begin{theorem}
\label{th:bmm-lb}
For every constant \(\varepsilon>0\), a
\((3-\varepsilon)\)-approximation algorithm for roundtrip APSP on
\(n\)-vertex directed unweighted graphs running in time \(T(n)\) implies a
Boolean matrix multiplication algorithm running in time
\(
 T(O(n))+O(n^2).
\)
\end{theorem}

The second reduction explains the asymmetric guarantee
\(
 2d(u\to v)+d(v\to u)
\)
obtained by our algorithms. It shows that the coefficient \(2\) of the
forward distance cannot be improved, even if the reverse distance is allowed
an arbitrary constant coefficient.

\begin{theorem}
\label{th:oneway-bmm-lb}
For all constants \(\varepsilon>0\) and \(C>0\), an algorithm that, on
\(n\)-vertex directed unweighted graphs, computes estimates satisfying
\[
 d(u\to v)
 \le
 \dist_{\mathcal D}(u,v)
 \le
 (2-\varepsilon)d(u\to v)+C\,d(v\to u)
\]
for all ordered pairs in time \(T(n)\) implies a Boolean matrix
multiplication algorithm running in time
\(
 T(O(n))+O(n^2).
\)
\end{theorem}

Thus, our lower bounds match the two main features of our stretch-\(3\)
algorithms: the final roundtrip stretch and the asymmetric one-way guarantee
used to obtain it.

\paragraph{Further upper bounds.}

The same techniques yield additional tradeoffs. First, by using a faster distance query and allowing a larger asymmetric error, we obtain a near-quadratic distance oracle.

\begin{theorem}
\label{thm:5-roundtrip-do}
There exists a distance oracle of expected size $O(n^{3/2})$ that can be constructed in expected time $\widetilde O(n^2)$ and answers each query $(v,u)$ in $O(\log n)$ time.
For every pair of vertices $v,u \in V$, the oracle returns a value $\dist_{\mathcal{D}}(v,u)$ satisfying
\[
d_G(v \to u)
\le
\dist_{\mathcal{D}}(v,u)
\le
3\,d_G(v \to u) + 2\,d_G(u \to v).
\]
\end{theorem}

Querying all pairs gives a $5$-approximation in time almost matching the $\Theta(n^2)$ output size.

\begin{corollary}
\label{cor:5-roundtrip-apsp}
For any directed graph with nonnegative edge weights, one can compute a $5$-roundtrip APSP approximation in expected time
\(
\widetilde O(n^2).
\)
\end{corollary}

Finally, when edge weights are bounded, some of the all-pairs aggregation steps in our algorithms can be accelerated using rectangular approximate min-plus products. This gives algebraic improvements over the combinatorial running times.

\begin{theorem}
\label{th:4eps-apsp}
Assume all edge weights lie in $[1,W]$.
There exists a randomized algorithm that computes a $(4+\varepsilon)$-roundtrip APSP approximation in expected time
\(
\widetilde O\!\left(n^{2.032}\log(nW)/\varepsilon\right).
\)
\end{theorem}

\begin{theorem}
\label{th:3eps-apsp}
Assume all edge weights lie in $[1,W]$.
There exists a randomized algorithm that computes a
$(3+\varepsilon)$-roundtrip APSP approximation in expected time
\(
\widetilde O\!\left(
(n^{2.037}+nm^{2/3})\log(nW)/\varepsilon
\right).
\)
\end{theorem}

\begin{table}[t]
\centering
\small
\caption{Summary of our upper and lower bounds for approximating all-pairs roundtrip distances in directed graphs. The algebraic bounds assume edge weights in $[1,W]$.}
\renewcommand{\arraystretch}{1.25}
\begin{tabular*}{\textwidth}{@{\extracolsep{\fill}}|c|c|c|c|}
\hline
\textbf{Approximation} & \textbf{Type} & \textbf{Running Time / Barrier} & \textbf{Reference} \\
\hline
\hline
$3$ & Combinatorial &
$\widetilde{O}(m\sqrt{n}+n^2)$ &
Corollary~\ref{cor:3-apsp} \\
\hline
$3$ & Combinatorial &
$\widetilde{O}(m^{1/3}n^{5/3} + n^2)$ &
Theorem~\ref{th:fast-3-apsp} \\
\hline
$4$ & Combinatorial &
$\widetilde{O}(n^{9/4})$ &
Theorem~\ref{th:4-apsp} \\
\hline
$5$ & Combinatorial &
$\widetilde{O}(n^2)$ &
Corollary~\ref{cor:5-roundtrip-apsp} \\
\hline
\hline
$(3+\varepsilon)$ & Algebraic &
$\widetilde{O}((n^{2.037}+nm^{2/3})\log(nW)/\varepsilon)$ &
Theorem~\ref{th:3eps-apsp} \\
\hline
$(4+\varepsilon)$ & Algebraic &
$\widetilde{O}(n^{2.032}\log(nW)/\varepsilon)$ &
Theorem~\ref{th:4eps-apsp} \\
\hline
\hline
$(3-\varepsilon)$ & Lower bound &
BMM reduction &
Theorem~\ref{th:bmm-lb} \\
\hline
$(2-\varepsilon)d(u \to v)+C \cdot d(v \to u)$ & Lower bound &
BMM reduction &
Theorem~\ref{th:oneway-bmm-lb} \\
\hline
\end{tabular*}
\label{tab:rt-apsp-summary}
\end{table}

\paragraph{Organization.}
Section~\ref{sec:prelim} introduces notation and key tools.
In Section~\ref{sec:3-apsp}, we present the first tool and use it to obtain a
$3$-roundtrip distance oracle and APSP algorithm.
In Section~\ref{sec:4-apsp}, we present the second tool and use it to obtain a
$4$-approximation algorithm.
In Section~\ref{sec:fast-3}, we combine these tools to obtain a faster
$3$-approximation APSP algorithm.
Section~\ref{sec:lb} presents the lower bounds.
Appendix~\ref{app:sec3} completes the analysis of
Section~\ref{sec:3-apsp}.
Appendix~\ref{app:5-apsp} gives the near-quadratic $5$-approximation algorithm,
and Appendix~\ref{app:fmm} presents the algebraic accelerations.

\section{Preliminaries}
\label{sec:prelim}

Let $G=(V,E, w)$ be a directed input graph with nonnegative edge weights $w: E \to \mathbb{R}_{\ge 0}$. For a directed graph $H$, let $V(H)$ and $E(H)$ denote the vertex and edge sets of $H$, respectively. For every $v,u \in V(H)$, denote by $d_H(u \to v)$ the (shortest path) distance from $u$ to $v$ in $H$. If there is no path from $u$ to $v$, we set $d_H(u \to v) = \infty$. Moreover, let $ u \lra v $ be the shortest closed walk containing $u$ and $v$, and the \textit{roundtrip distance} between $u$ and $v$ as $d_H(u \leftrightarrows v):= d_H(u \to v) + d_H(v \to u) $ (which is the length of that shortest closed walk). When the underlying graph is clear, we drop the subscript and write $d$ instead of $d_G$.
Let $N_G(v)$ be the \textbf{outgoing} neighbors of $v \in V$.

\paragraph{Roundtrip distance oracles.}
A \emph{(roundtrip) approximate distance oracle} for a directed graph $G=(V,E,w)$ is a data structure that, after preprocessing $G$, can answer queries for any pair of vertices $u,v\in V$ with an estimate $\dist_\D(u \lra v)$ of the roundtrip distance $d(u \lra v)$.
An oracle has \emph{stretch} $\alpha \ge 1$ if for all $u,v \in V$,
\[
d(u \lra v) \;\le\; \dist_\D(u \lra v) \;\le\; \alpha \cdot d(u \lra v).
\]
The performance of an oracle is measured by its preprocessing time, space usage, and query time.

\paragraph{Stored-distance interface.}
Our distance oracles maintain a \emph{sparse} set of stored directed distances.
Formally, we view the stored data as a partial function
\[
\delta:\ \Dom(\delta)\subseteq V\times V \ \to\ \mathbb{R}_{\ge 0},
\]
where $(x,y)\in\Dom(\delta)$ (the domain of $\delta$) iff the construction explicitly stored a value for this pair.
For notational convenience, we extend $\delta$ to all pairs by defining
\[
\delta(x,y)=\infty \quad \text{for } (x,y)\notin \Dom(\delta), \qquad \text{and } \, \, \delta(x,x) = 0. 
\]
We assume standard expected $O(1)$ dictionary access.

\paragraph{Degree balancing.}
We assume throughout that
\(
|N(v)|=O(m/n) \) for every \( v\in V .\)
This is without loss of generality: the standard degree-reduction
transformation of~\cite{roundtrip_spanners19} produces, in \(O(m)\) time, a graph
with \(O(n)\) vertices and \(O(m)\) edges, maximum outdegree \(O(m/n)\),
and the same distances between the original vertices.

\section{The Pivot Machinery and the First $3$-APSP Approximation}
\label{sec:3-apsp}

This section is devoted to the proof of Theorem~\ref{th:3do} and its application to APSP, Corollary~\ref{cor:3-apsp}. We first isolate the main pivot mechanism, which is based on an observation of Chechik and Lifshitz~\cite{CL21}, and then show how to turn it into a distance oracle.

This should be viewed as a baseline result. The pivot mechanism is based on known ideas, but we use it here to obtain the distance-oracle needed for roundtrip APSP. This also introduces the main primitive that will later be combined to obtain a faster algorithm. Since the analysis closely follows the known arguments, its full details are deferred to Appendix~\ref{app:sec3}.

\subsection{The Pivot Machinery}

The starting point is the following question. Suppose that all distances to and from a vertex \(r\) have already been computed. For which pairs \((v,u)\) can \(r\) be used to estimate \(d(v\to u)\), and what does this imply for the corresponding roundtrip distances?

Chechik and Lifshitz~\cite{CL21} observed that it suffices to test the
following comparison. If, for vertices \(v,u,r\),
\begin{equation}\label{eq:key}
 2d(v \to r)+d(r \to u)
 \le
 2d(v \to u)+d(u \to r),
\end{equation}
then \(r\) provides the guarantees stated in the following lemma.

\begin{lemma}[Chechik--Lifshitz~\cite{CL21}]
\label{lem:key-observation}
Let \(G=(V,E,w)\) be a directed graph with nonnegative edge weights, and let
\(v,u,r\in V\). If~\eqref{eq:key} holds, then
\[
 d(v\to r)+d(r\to u)
 \le
 2d(v \to u)+d(u\to v).
\]
Moreover,
\[
 d(v\lra r),\ d(u\lra r)
 \le
 2d(v\lra u).
\]
\end{lemma}

\begin{figure}[h]
 \centering
 \begin{tikzpicture}

 \node[circle,draw,inner sep=2.5pt,label=below:$v$] (v) at (0,-2) {};
 \node[circle,draw,inner sep=2.5pt,label=right:$u$] (u) at (1.5,0) {};
 \node[circle,draw,inner sep=2.5pt,label=left:$r$] (r) at (-1.5,0) {};

 \draw[directed, purple] (v) to (u);
 \draw[line width=4pt, green, opacity=0.3, bend left=45] (u) to (v);
 \draw[directed, purple, bend left=45] (u) to (v);
 \draw[directed] (v) to (r);
 \draw[line width=4pt, green, opacity=0.3] (v) to (r);
 \draw[directed, bend left=30] (u) to (r);
 \draw[directed, bend left=30] (r) to (u);
 \draw[line width=4pt, green, opacity=0.3, bend left=30] (r) to (u);

 \end{tikzpicture}
 \caption{An illustration of Lemma~\ref{lem:key-observation}. The directed
 cycle \(u\to v\to u\) (red) can be \(2\)-approximated by the directed
 cycle \(u\to v\to r\to u\) (green) when~\eqref{eq:key} holds.}
 \label{fig:key-observation}
\end{figure}
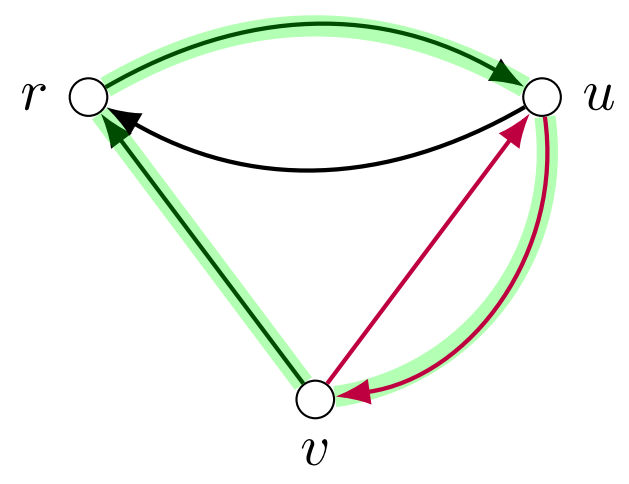

Figure~\ref{fig:key-observation} illustrates the lemma. Intuitively, if
\(r\) passes the comparison~\eqref{eq:key}, then the cycle through \(v,u\)
can be replaced, up to a factor \(2\), by a cycle that goes through the
pivot \(r\).

The important algorithmic point is not only that a pivot can provide a good approximation for one pair, but also that a random pivot can provide good approximations for many pairs simultaneously. Fix a vertex \(v\), and consider some active set \(B(v)\) of vertices whose distance from \(v\) do not have a good enough approximation. For every two vertices \(u,r\in B(v)\), at least one of the two comparisons
\[
 2d(v\to r)+d(r\to u)
 \le
 2d(v\to u)+d(u\to r)
\]
or
\[
 2d(v\to u)+d(u\to r)
 \le
 2d(v\to r)+d(r\to u)
\]
holds. Thus, for the fixed source \(v\), these comparisons orient every
pair in \(B(v)\). If the first inequality holds, we say that \(r\) \textbf{eliminates} \(u\) with respect to \(v\).

This creates a tournament on the active set \(B(v)\). Consequently, a
uniformly random pivot \(r\in B(v)\) eliminates half of the remaining
vertices in expectation. The oracle below implements this idea implicitly: it samples pivots at increasing rates, stores distances to and from them, and then runs a truncated search from every vertex \(v\), continuing only through vertices that have not been eliminated by an already chosen pivot.

We now show how to turn this pivot-elimination mechanism into a
\(3\)-roundtrip distance oracle. Querying the oracle on all pairs then gives Corollary~\ref{cor:3-apsp} in total expected time
\(\widetilde O(m\sqrt n+n^2)\).

\subsection{Construction Overview}

The oracle preprocessing proceeds in two phases.

\paragraph{Phase 1: Sample the pivots.}
Set
\[
\Delta := \left\lceil \log_{3/2} \sqrt{n} \right\rceil,
\qquad
\alpha := (\sqrt{n})^{1/\Delta}.
\]
Note that $\alpha \in [5/4,3/2]$ for $\sqrt{n}\ge 2$ and $\alpha^\Delta=\sqrt{n}$.
For each vertex $v\in V$, the oracle maintains a pivot set $P(v)$, initially empty.

For iterations $i=0,1,\ldots,\Delta-1$, the algorithm samples a set $S_i \subseteq V$ by including each vertex independently with probability $\alpha^i/n$, so that $\mathbb{E}[|S_i|]=\alpha^i$.
For each sampled witness $ s \in S_i$, it runs both outgoing and incoming Dijkstra searches rooted at $s$ in the original graph $G$. The oracle stores these distances by setting
$\delta(s,v)\gets d(s\to v)$ and $\delta(v,s)\gets d(v\to s)$ for all $v\in V$ and $s\in S_i$.

After processing the witnesses in $S_i$, it updates \textbf{pivots}: for each $v\in V$, if there exists a witness $s\in S_i$ such that for every $p \in P(v)$,
\[
2d(v \to s) + d(s \to p) \;<\; 2d(v \to p) + d(p \to s),
\]
then it adds one such $s$ to $P(v)$ (choosing one uniformly at random if multiple exist). Note that all those distances are stored via $\delta$. This ensures that by the end of Phase~1 we have $|P(v)|=O(\Delta) = O(\log n)$ for every $v$.

\paragraph{Phase 2: truncated outgoing searches.} After $\Delta$ iterations, for each vertex $v \in V$, the algorithm runs an outgoing Dijkstra search rooted at $v$, but \emph{truncates} the search using the current pivot set $P(v)$.

Concretely, when a vertex $u$ is extracted from the priority queue, let $\widehat d(v \to u)$ denote its current distance label. The algorithm scans the outgoing edges of $u$ only if, for every pivot $p \in P(v)$,
\[
2\widehat d(v \to u) + d(u \to p) \;<\; 2d(v \to p) + d(p \to u).
\]
Let $T(v)$ denote the resulting truncated search tree rooted at $v$. 

For each vertex $u \in T(v)$, the oracle stores the distance found by the search by setting \( \delta(v,u) \gets d_{T(v)}(v \to u), \) where $d_{T(v)}(v \to u)$ denotes the length of the path from $v$ to $u$ in $T(v)$.

In the analysis, we show that $\mathbb{E}[|T(v)|]=O(\sqrt{n})$ for every $v\in V$, implying that the oracle stores only $O(n^{3/2})$ total space in expectation. The construction is given in Algorithm~\ref{alg:3do-construct}.

\begin{algorithm}[H]
\SetAlgoNoEnd
\caption{Construction of a $3$-roundtrip distance oracle}
\label{alg:3do-construct}
\DontPrintSemicolon

\textbf{Input:} Directed graph $G=(V,E,w)$ with $w\ge 0$ \\
\textbf{Output:} Distance oracle $\mathcal{D}$

\medskip
Set \(
\Delta \gets \left\lceil \log_{3/2} \sqrt{n} \right\rceil,
\qquad
\alpha \gets (\sqrt{n})^{1/\Delta}.
\)

Set $P(v) \gets \emptyset$ for each $v \in V$.

\medskip
\textbf{Phase 1: Sample the pivots}

\For{$i \gets 0$ \KwTo $\Delta-1$}{
 Sample $S_i \subseteq V$ by including each vertex independently
 with probability $\alpha^i / n$.\;
 
 Run outgoing and incoming Dijkstra from each $s \in S_i$ in $G$ \;
 Set $\delta(s,v) \gets d(s \to v), \, \delta(v,s) \gets d(v \to s)$ for every $v \in V$ and $s \in S_i$\;

 \ForEach{$v \in V$}{
 \If{$\exists s \in S_i$ such that
 \[
 \forall p \in P(v):\;
 2d(v \to s) + d(s \to p)
 < 2d(v \to p) + d(p \to s)
 \]
 }{
 Add one such $s$ to $P(v)$.\;
 }
 }
}

\medskip
\textbf{Phase 2: Truncated searches}

\ForEach{$v \in V$}{
 Run outgoing Dijkstra from $v$. When a vertex $u$ with current distance label $\widehat d(v \to u)$ is extracted, scan its outgoing edges only if
 \[
 \forall p \in P(v):\; 2\widehat d(v \to u) + d(u \to p) < 2d(v \to p) + d(p \to u)
 \]
 Let the resulting truncated tree be $T(v)$.\;
 Set $\delta(v,u) \gets d_{T(v)}(v \to u)$ for all $u \in T(v)$\;
 Add $v$ to $P(v)$\;
}

\medskip
\end{algorithm}

\paragraph{Distance query.}
Given $(v,u)$, the oracle returns the minimum of distances from $v$ to $u$ via a pivot $p\in P(v)$ using the stored information function $\delta$:
\[
\dist_{\mathcal{D}}(v,u)
 \;=\;
 \min_{p\in P(v)} \bigl\{ \delta(v,p) + \delta(p,u) \bigr\}.
\]
Since $|P(v)|=O(\log n)$, the query time is $O(\log n)$.
We summarize the query in Algorithm~\ref{alg:3do-query}.

\begin{algorithm}[H]
\caption{Query procedure for $\dist_{\mathcal{D}}(v,u)$}
\label{alg:3do-query}
\DontPrintSemicolon

\textbf{Input:} Vertices $v,u \in V$ \\
\textbf{Output:} Approximate distance $\dist_{\mathcal{D}}(v,u)$

\Return
\(
 \min_{p\in P(v)} \bigl\{ \delta(v,p) + \delta(p,u) \bigr\}.
\)\;
\end{algorithm}

The analysis of the algorithm with its full proofs appears in Appendix~\ref{app:sec3}.

\section{Sparsification Using Centers and the $4$-APSP Approximation}
\label{sec:4-apsp}

In this section, we prove Theorem~\ref{th:4-apsp}. The main ingredient is a sparsification idea of Chechik and Lifshitz~\cite{CL21}, originally used for constructing $4$-roundtrip spanners. We adapt it to the all-pairs setting by storing distance estimates before deleting edges.

\subsection{Sparsification Using Centers}

We now turn to the second ingredient of our framework. As in the pivot machinery, suppose that we have already computed shortest paths from and to a vertex \(r\). Consider an edge \((a,b)\in E\) such that \(r\) eliminates \(b\) with respect to \(a\), namely, such that~\eqref{eq:key} holds.

This means that \(a \to r \to b\) is a good replacement for the local edge \((a,b)\), but the consequence is stronger. If a shortest path from \(v\) to \(u\) uses the edge \((a,b)\), then the path $v \to r \to u$ produces a good approximation for the distance $d(v \to u)$. In particular, if \((a,b)\) lies on the shortest roundtrip
\(v\lra u\), then the cycle 
\[
 v\to a\to r\to b\to u\to v
\]
has length at most \(2d(v\lra u)\).

Our sparsification procedure uses this observation as follows. We sample centers and use them to eliminate outgoing edges. Before deleting such an edge, we store roundtrip estimates through the center responsible for its deletion. In the setting above, when \(r\) is responsible for the deletion of an edge on a shortest path from \(v\) to \(u\), we obtain
\begin{align}
d(v \to r) + d(r \to u) \, \le & \, 2d(v \to u) + d(u \to v), \label{eq:4-approx2} \quad \text{so }
\\
d(v\lra r),d(u\lra r) \, \le & \, 2d(v\lra u). \label{eq:4-approx}
\end{align}
 
\begin{figure}[h]
 \centering
 \begin{tikzpicture}

 \node[circle,draw,inner sep=2.5pt,label=below left:$a$] (a) at (0,-1) {};
 \node[circle,draw,inner sep=2.5pt,label=below:$b$] (b) at (0.75,0) {};
 \node[circle,draw,inner sep=2.5pt,label=left:$r$] (s) at (-0.75,0) {};

 \node[circle,draw,inner sep=2.5pt,label=below:$v$] (v) at (2,-2) {};
 \node[circle,draw,inner sep=2.5pt,label=right:$u$] (u) at (2,0.3) {};

 \draw[directed] (a) to (s);
 \draw[line width=4pt, green, opacity=0.3] (a) to (s);
 \draw[directed, purple, dashed] (a) to (b);
 \draw[directed, bend left=30] (s) to (b);
 \draw[bend left=30,line width=4pt, green, opacity=0.3] (s) to (b);
 \draw[directed, bend left=30] (b) to (s);

 \draw[directed, purple, bend left=20] (b) to (u);
 \draw[line width=4pt, green, opacity=0.3, bend left=20] (b) to (u);
 \draw[directed, purple, bend left=50] (u) to (v);
 \draw[line width=4pt, green, opacity=0.3, bend left=50] (u) to (v);
 \draw[directed, purple, bend left=40] (v) to (a);
 \draw[line width=4pt, green, opacity=0.3, bend left=40] (v) to (a);

 \end{tikzpicture}
 \caption{An illustration of Inequality~\eqref{eq:4-approx}. The cycle
 \(u\to v\to u\) (red) can be \(2\)-approximated by the cycle
 \(v\to a\to r\to b\to u\to v\) (green) when~\eqref{eq:key} holds.}
 \label{fig:4-stretch}
\end{figure}
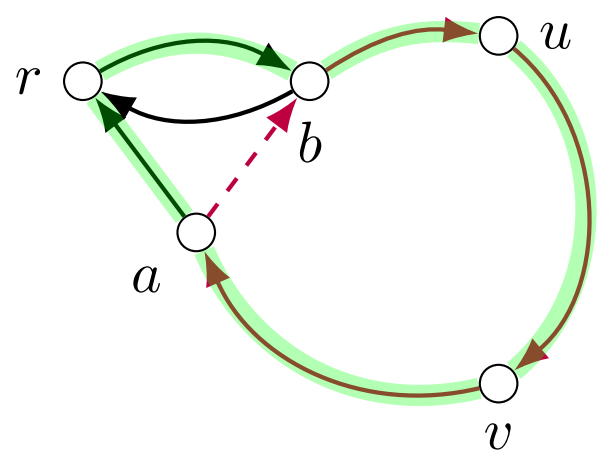

\medskip

We use the sparsification procedure to construct a sparse subgraph
\(H\subseteq G\), together with explicit estimates for all pairs whose
shortest roundtrip is destroyed by the sparsification. The key guarantee is
the following.

\begin{lemma}\label{lem:roundtrip-oracle-comb}
Let \(G=(V,E,w)\) be a directed graph with nonnegative edge weights, and let
\(\beta\in[0,1]\). There is a randomized algorithm that, in expected time
\(
 \widetilde O(n^{2+\beta}),
\)
computes:
\begin{enumerate}
 \item a subgraph \(H\subseteq G\) of expected size \(O(n^{2-\beta})\), and
 \item an estimate \(\delta(v\lra u)\) for every pair \(v,u\in V\),
\end{enumerate}
such that the following holds for every pair \(v,u\in V\):
\begin{enumerate}
 \item \(d_G(v\lra u)\le \delta(v\lra u)\), and
 \item if the shortest roundtrip \(v\lra u\) is not fully contained in
 \(H\), then
 \[
 \delta(v\lra u)\le 4d_G(v\lra u).
 \]
\end{enumerate}
\end{lemma}

Our \(4\)-approximation algorithm now follows directly. Apply
Lemma~\ref{lem:roundtrip-oracle-comb} with parameter \(\beta\), obtaining a
subgraph \(H\) with \(O(n^{2-\beta})\) expected edges and estimates
\(\delta\). Then run the \(3\)-approximation algorithm of
Corollary~\ref{cor:3-apsp} on \(H\). For every pair \(v,u\), if the shortest
roundtrip \(v\lra u\) is destroyed by the sparsification, the estimate
\(\delta(v\lra u)\) is a \(4\)-approximation. Otherwise, the shortest
roundtrip remains in \(H\), and the \(3\)-approximation on \(H\) gives a
valid, hence also \(4\)-approximate, estimate in \(G\). Taking the minimum of
the two estimates therefore gives a \(4\)-roundtrip APSP approximation.

The running time is
\[
 \widetilde O\left(n^{2+\beta}+|E(H)|\sqrt n+n^2\right)
 =
 \widetilde O\left(n^{2+\beta}+n^{5/2-\beta}\right).
\]
Choosing \(\beta=1/4\) balances the two terms and gives
\(
 \widetilde O(n^{9/4}),
\)
proving Theorem~\ref{th:4-apsp}.

\subsection{Construction Overview}

We now describe the algorithm achieving Lemma~\ref{lem:roundtrip-oracle-comb}.
The construction closely follows the $4$-roundtrip spanner algorithm of
Chechik and Lifshitz~\cite{CL21}, with three major differences:
(i) the sparsification process is stopped early, once the remaining graph has
at most $n^{2-\beta}$ edges;
(ii) in every iteration, roundtrip distances are explicitly aggregated via the
sampled centers in order to obtain distance estimates for all vertex pairs; and
(iii) the sampling technique of Harbuzova et al.~\cite{HarbuzovaJinVWXu24} is incorporated.

The algorithm maintains a sequence of graphs $G_0, G_1, \dots$, obtained by
iteratively deleting edges, together with a table $\delta$ of roundtrip
distance estimates. Initially, let $G_0 := G$. The algorithm sets
\[
\Delta := \left\lceil \beta \log_{3/2} n \right\rceil
\quad\text{iterations, and}\quad
\alpha := n^{\beta / \Delta}.
\]
Note that $\alpha \in [5/4,3/2]$ for $n^\beta \ge 2$, and $\alpha^\Delta = n^\beta$.
For iterations $i=0,1,\ldots,\Delta-1$, the following steps are executed.

\paragraph{Sampling centers.}
In iteration $i$, a set $S_i \subseteq V$ is sampled by including each vertex
independently with probability $\alpha^i / n$.
Hence $\mathbb{E}[|S_i|] = \alpha^i$.

\paragraph{Computing distances and aggregating estimates.}
Let $G_i$ denote the current graph at the beginning of iteration $i$.
For every $s \in S_i$, outgoing and incoming Dijkstra searches are run in $G_i$,
yielding roundtrip distances $d_i(s \lra \cdot)$.

For every pair of vertices $v,u \in V$, the algorithm updates the global estimates as
\[
 \delta(v \lra u)
 \;:=\;
 \min_{s\in S_i}\bigl\{\delta(v \lra u),\, d_i(v \lra s) + d_i(s \lra u)\bigr\}.
\]

\paragraph{Edge pruning.}
For each vertex $v \in V$, let $N_i(v)$ denote its outgoing neighbors in $G_i$.
If $N_i(v) \cap S_i \neq \emptyset$, a center
$c_i(v) \in N_i(v) \cap S_i$ is chosen uniformly at random.

The algorithm then inspects every edge $(v,u) \in E(G_i)$.
If $v$ has an assigned center $c_i(v)$ and
\begin{equation}
 2\,d_i(v \to c_i(v)) + d_i(c_i(v) \to u)
 \;\leq\;
 2\,w(v,u) + d_i(u \to c_i(v)),
\label{eq:4-rule}
\end{equation}
the edge $(v,u)$ is deleted.
Let $G_{i+1}$ denote the resulting graph.

Note that this is a slight modification of the original deletion rule: we replace \(d_i(v \to u)\) by the current edge weight \(w(v,u)\). This is needed because computing \(d_i(v \to u)\) for every edge would be too expensive. The change is harmless, since always \(d_i(v \to u)\le w(v,u)\). If \(d_i(v \to u)<w(v,u)\), then the edge \((v,u)\) cannot belong to any shortest roundtrip cycle in \(G_i\) and can be removed.

After all iterations are completed, the algorithm returns $(G_\Delta, \delta)$.
The pseudo-code is given in Algorithm~\ref{alg:4-apsp}.

\begin{algorithm}[H]
\SetAlgoNoEnd
\caption{Roundtrip Sparsification with $4$-Approximation estimates}
\label{alg:4-apsp}
\DontPrintSemicolon

\KwIn{Directed graph $G=(V,E,w)$ with $w \ge 0$, parameter $\beta\in(0,1)$.}
\KwOut{Subgraph $H \subseteq G$ and estimates $\delta( \cdot \lra \cdot )$.}

Set $G_0 \gets G$; and $\Delta \gets \left\lceil \beta \log_{3/2} n \right\rceil$; and $\alpha \gets n^{\beta/\Delta}$; \;

\For{$i \gets 0$ \KwTo $\Delta-1$}{
 Sample $S_i \subseteq V$ by including each $v \in V$ independently with probability $\alpha^i / n$; \;

 Run outgoing and incoming Dijkstra from every $s \in S_i$ in $G_i$ to obtain $d_i(s \lra \cdot)$; \;
 \ForEach{$v,u \in V$}{
 Set $\delta(v \lra u) \gets \min_{s \in S_i}\{\delta(v \lra u),\, d_i(v \lra s) + d_i(s \lra u)\}$; \;
 }

 $G_{i+1} \gets G_i$; \;

 \ForEach{$v \in V$ such that $N_i(v) \cap S_i \neq \emptyset$}{
 pick a random $c_i(v) \in N_i(v) \cap S_i$; \;
 \ForEach{$u \in N_i(v)$}{
 \If{$2 d_i(v \to c_i(v)) + d_i(c_i(v) \to u) \le 2w(v,u) + d_i(u \to c_i(v))$}{
 delete edge $(v,u)$ from $G_{i+1}$; \;
 }
 }
 }
}

\Return{$(G_{\Delta},\delta)$}\;
\end{algorithm}

\subsection{Analysis}

We start by analyzing the stretch.
\begin{lemma}\label{lem:roundtrip-stretch-general}
Let $v,u\in V$. If the shortest roundtrip $v \lra u$ is not fully contained in $G_\Delta$, then
\[
\delta(v \lra u) \le 4\, d(v \lra u).
\]
\end{lemma}

\begin{proof}
Let $C$ be a shortest roundtrip between $v$ and $u$ in $G$, and let $i$ be the last iteration such
that $C \subseteq G_i$.
Then there exists an edge $e=(a,b)$ on $C$ such that $e\in G_i \setminus G_{i+1}$. See Figure \ref{fig:4-stretch}.

The edge $(a,b)$ was deleted in iteration $i$ due to a center $c:=c_i(a)$ satisfying
\begin{equation}
 \label{eq:4assumption}
 2\,d_i(a \to c) + d_i(c \to b)
\;\le\;
2\,w(a,b) + d_i(b \to c).
\end{equation}
Since $(a,b)$ lies on the shortest roundtrip $C$, we have $w(a,b)=d_i(a\to b)$.
Substituting this into the deletion inequality and using the triangle inequality
$d_i(b\to c)\le d_i(b\to a)+d_i(a\to c)$ yields
\begin{equation}
\label{eq:lem43}
 d_i(a \to c) + d_i(c \to b)
\;\le\;
2\,d_i(a \to b) + d_i(b \to a)
\;\le\;
d_i(a \to b) + d_i(v \lra u),
\end{equation}
where the last inequality holds since $(a,b)$ lies on the roundtrip $C$.

Now fix any vertex $z$ on the roundtrip $v \lra u$.
Then
\begin{align*}
d_i(z \lra c)
&\le d_i(z \to a) + d_i(a \to c) + d_i(c \to b) + d_i(b \to z) && \text{by~\eqref{eq:lem43}}\\
&\le d_i(z \to a) + d_i(a \to b) + d_i(b \to z) + d_i(v \lra u) \\
&\le 2\,d_i(v \lra u),
\end{align*}
where the last inequality follows from the fact that $z \to a \to b \to z$ is exactly $C$. Applying this bound with $z=v$ and $z=u$ implies
\begin{equation}
d_i(v \lra c),\ d_i(c \lra u) \;\le\; 2\,d_i(v \lra u). 
\end{equation}

Therefore, in iteration $i$ the algorithm sets
\[
\delta(v \lra u)
\;\le\;
d_i(v \lra c) + d_i(c \lra u)
\;\le\;
4\, d_i(v \lra u),
\]
as claimed.

\end{proof}

We next analyze the size of the remaining graph.
Bounding the number of edges after each iteration follows essentially the same argument as in
Claim~\ref{clm:Bi} (and Lemma~4.3 in~\cite{HarbuzovaJinVWXu24}), so we do not repeat the full proof.
The only difference is that Claim~\ref{clm:Bi} bounds, for each vertex $v\in V$, the size of the set
$B_i(v)$, whereas here we bound the number of outgoing neighbors $N_i(v)$ of $v$ in iteration $i$.

\begin{lemma}\label{lem:4do-edge-bound}
For every iteration $i$ and every $v \in V$,
\[
\mathbb{E}[|N_i(v)|]\ \le\ 3\cdot \frac{n}{\alpha^i}, \quad \text{therefore } \, \mathbb{E}\bigl[|E(G_i)|\bigr] \;\le\; \frac{3n^2}{\alpha^i}.
\]
In particular, since $\alpha^\Delta = n^\beta$, it follows that
\[
\mathbb{E}\bigl[|E(G_\Delta)|\bigr] \;=\; O(n^{2-\beta}).
\]
\end{lemma}

Lastly, we bound the expected running time of the sparsification procedure.

\begin{claim}\label{clm:running-time-sparsify}
The sparsification procedure runs in expected time
\(
\tilde{O}(n^{2+\beta}).
\)
\end{claim}

\begin{proof}
The running time has two components: distance aggregation and edge pruning.

\paragraph{Aggregation.}
In iteration $i$, the algorithm enumerates all centers $s\in S_i$ and all ordered pairs
$(v,u)\in V\times V$ to update $\delta(v \lra u)$, which takes $O(n^2|S_i|)$ time.
Since $\e[|S_i|]=\alpha^i$ and $\sum_{i<\Delta}\alpha^i=O(n^\beta)$, the total expected time
spent on aggregation is $O(n^{2+\beta})$.

\paragraph{Edge pruning.}
In iteration $i$, edge pruning is dominated by, for each vertex $v$, identifying a center in
$N_i(v)\cap S_i$ and testing the pruning condition on all outgoing edges of $v$.
As $S_i$ is independent of $G_i$, we have
\[
\e\bigl[|E(G_i)|\cdot |S_i|\bigr]
=
\e\bigl[|E(G_i)|\bigr]\cdot \e\bigl[|S_i|\bigr]
=
O(n^2),
\]
using Lemma~\ref{lem:4do-edge-bound}.
Summing over all $\Delta=O(\log n)$ iterations yields a total expected pruning time of
$O(n^2\log n)$.

Combining the two parts gives an overall expected running time of
$\tilde{O}(n^{2+\beta})$.
\end{proof}

\section{Combining Pivots and Centers for a Faster $3$-Approximation}
\label{sec:fast-3}

In this section we prove Theorem~\ref{th:fast-3-apsp}. The algorithm combines the two mechanisms developed in the previous sections: the pivot machinery of the \(3\)-APSP oracle, and the center-based sparsification used for the \(4\)-APSP. The goal is to retain the same one-way guarantee as in Theorem~\ref{th:3do}, namely
\[
 \delta(v,u)
 \le
 2d(v\to u)+d(u\to v)
\]
for every ordered pair \((v,u)\), while running the expensive truncated-search phase on a sparser graph.

Recall the bottleneck in the basic pivot-based algorithm. If the sampling process reaches density about \(q/n\), then the relevant truncated Dijkstra tree for a vertex $v$ has expected size roughly
\(
 |T(v)| \le n/q .
\)
The searches from the sampled pivots cost \(\widetilde O(mq)\). The second phase is the bottleneck: even though the truncated search from a vertex visits only about \(|T(v)|\) vertices, it is still performed in the original graph. Under the degree-balancing assumption, the total cost of these searches is roughly
\[
 n\cdot \widetilde O\!\left(\frac{m}{n}|T(v)|\right)
 =
 \widetilde O(m|T(v)|).
\]
Balancing the two terms gives the choice \(q=\sqrt n\), and hence the running time \(\widetilde O(m\sqrt n+n^2)\).

Our improvement is to stop the sampling earlier, at \(q=n^{1/3}\), and to sparsify the graph \textbf{only once} between the sampling process and the truncated searches. With this choice, the truncated balls have expected size
\(
 \e[|T(v)|] \le n/q = n^{2/3}.
\)
If the sparsification reduces the residual outdegree of every vertex to \(\widetilde O(n/q)\), then a truncated search that visits \(|T(v)|\) vertices scans only \(\widetilde O(n/q \cdot |T(v)|)\) edges. Thus all final searches together cost
\[
 n \cdot \widetilde O \! \left( \frac{n}{q}|T(v)| \right) =\widetilde O \! \left( \frac{n^3}{q^2} \right) = \widetilde O(n^{7/3}),
\]
instead of \(\widetilde O(m|T(v)|)\).

The main question is how to use the sparsification procedure, which previously yielded a \(4\)-approximation, to obtain a \(3\)-approximation. Recall what can go wrong. The procedure may delete an edge on a shortest \(v\to u\) path and thereby obtain a good estimate for this direction (see~\eqref{eq:4-approx2}), while the shortest \(u\to v\) path remains intact and no good estimate for it is stored. If we wait until an edge on $u \to v$ is later deleted, the original \(v\to u\) path may already have been destroyed, and the resulting estimate need not be good. This is why the \(4\)-APSP algorithm stores a
roundtrip estimate for \(v\lra u\) as soon as the first edge of the roundtrip is deleted.

We also cannot run the \(3\)-APSP oracle on a sparsified graph. Instead, in the \textit{first phase}, for each vertex \(v\in V\), we construct two sets in parallel: a pivot set \(P(v)\subseteq V\), as in the basic \(3\)-approximation oracle, and a center set \(C(v)\subseteq N(v)\).

The center set is used in the \textit{second phase} to sparsify the graph in one shot. For every vertex \(v\), the algorithm inspects the outgoing edges of \(v\) and deletes every edge that has a good replacement through one of the centers in \(C(v)\). After this sparsification step, every shortest path from \(v\) to \(u\) satisfies one of two alternatives: either the path remains entirely in the sparsified graph \(H\), or some edge \((a,b)\) on the path $v \to u$ was deleted. In the latter case, the path $v \to c \to u$, where \(c\in C(a)\) is the center responsible for deleting \((a,b)\), serves as an estimate for the distance $d(v \to u)$.

The pivot set has the same role as in the \(3\)-APSP oracle,
except that the final truncated Dijkstra search is run on the sparsified graph \(H\). Thus, in the \textit{third phase}, if a shortest \(v\to u\) path remains in \(H\), then either the
truncated search from \(v\) reaches \(u\), in which case the algorithm stores the exact distance \(d(v\to u)\), or the search is stopped before reaching \(u\). In the latter case, as in the first \(3\)-APSP algorithm, some pivot \(p\in P(v)\) provides a good approximation for \(d(v \to u)\).

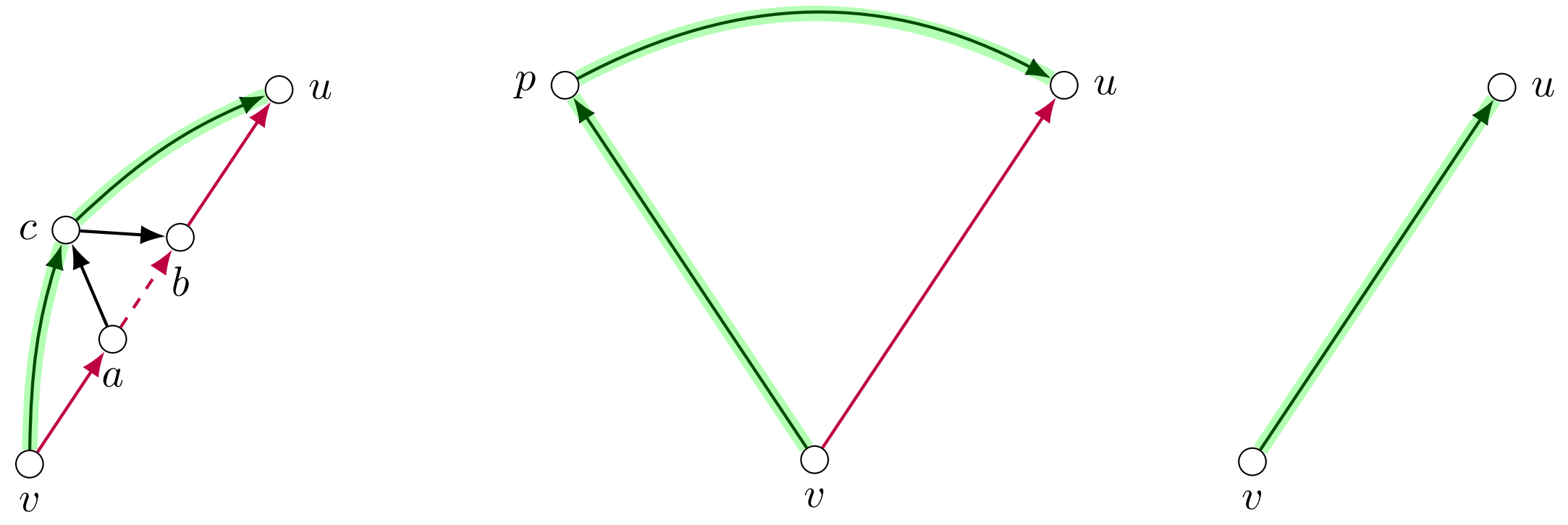
\begin{figure}[t]
\centering

\begin{subfigure}[t]{0.32\textwidth}
\centering
\begin{tikzpicture}[scale=0.95]

 \node[circle,draw,inner sep=2.5pt,label=below:$v$] (v) at (0,-1.5) {};
\node[circle,draw,inner sep=2.5pt,label=below:$a$] (a) at (0.8,-0.3) {};
\node[circle,draw,inner sep=2.5pt,label=below:$b$] (b) at (1.45,0.68) {};
\node[circle,draw,inner sep=2.5pt,label=right:$u$] (u) at (2.4,2.1) {};
\node[circle,draw,inner sep=2.5pt,label=left:$c$] (c) at (0.35,0.75) {};

 \draw[directed, purple] (v) to (a);
 \draw[directed, purple, dashed] (a) to (b);
 \draw[directed, purple] (b) to (u);

 \draw[directed] (a) to (c);
 \draw[directed] (c) to (b);
 
 \draw[directed, bend left=8] (v) to (c);
 \draw[line width=4pt, green, opacity=0.3, bend left=8] (v) to (c);

 \draw[directed, bend left=10] (c) to (u);
 \draw[line width=4pt, green, opacity=0.3, bend left=10] (c) to (u);

 \node at (1.2,-2.8) {\(\pi \not\subseteq H\)};

\end{tikzpicture}
\caption{The shortest path is destroyed in the sparsification.}
\end{subfigure}
\hfill
\begin{subfigure}[t]{0.32\textwidth}
\centering
\begin{tikzpicture}[scale=0.95]

 \node[circle,draw,inner sep=2.5pt,label=below:$v$] (v) at (0,-1.5) {};
\node[circle,draw,inner sep=2.5pt,label=right:$u$] (u) at (2.4,2.1) {};
\node[circle,draw,inner sep=2.5pt,label=left:$p$] (p) at (-2.4,2.1) {};

 \draw[directed, purple] (v) to (u);

 \draw[directed] (v) to (p);
 \draw[line width=4pt, green, opacity=0.3] (v) to (p);

 \draw[directed, bend left=28] (p) to (u);
 \draw[line width=4pt, green, opacity=0.3, bend left=28] (p) to (u);

 \node at (0,-2.8) {\(\pi \subseteq H\), but \(\pi \nsubseteq T_H(v)\)};

\end{tikzpicture}
\caption{The path survives, but the truncated search stops early.}
\end{subfigure}
\hfill
\begin{subfigure}[t]{0.32\textwidth}
\centering
\begin{tikzpicture}[scale=0.95]

 \node[circle,draw,inner sep=2.5pt,label=below:$v$] (v) at (0,-1.5) {};
\node[circle,draw,inner sep=2.5pt,label=right:$u$] (u) at (2.4,2.1) {};

\draw[directed] (v) to (u);
 \draw[line width=4pt, green, opacity=0.3] (v) to (u);

 \node at (2.1,-2.8) {\(\pi \subseteq T_H(v)\)};

\end{tikzpicture}
\caption{The path survives and is fully explored.}
\end{subfigure}

\caption{The three cases in the proof of the fast \(3\)-APSP algorithm. 
Let \(\pi = v \to u \). If \(\pi\) is not fully contained in \(H\), then some deleted edge \((a,b)\) on \(\pi\) is estimated by a center \(c\); the local replacement \(a\to c\to b\) implies that the algorithm may use the path \(v\to c\to u\) (left). 
If \(\pi\subseteq H\) but \(u\notin T_H(v)\), then a pivot \(p\in P(v)\) estimates the pair \((v,u)\) (middle). 
If \(u\in T_H(v)\), then the truncated search stores the exact distance \(d(v\to u)\) (right).}
\label{fig:fast-3-cases}
\end{figure}

\subsection{Construction Overview}

We now describe the algorithm proving Theorem~\ref{th:fast-3-apsp}. The algorithm maintains a subgraph \(H\), initially \(H:=G\), and a table \(\delta\) of directed distance estimates. For each vertex \(v\in V\), it also maintains two sets: a pivot set \(P(v)\subseteq V\), and a center set \(C(v)\subseteq N(v)\). The pivot set will be used to truncate the final searches, as in the basic \(3\)-APSP oracle. The center set will be used to sparsify the outgoing edges of \(v\). Set
\[
 \Delta := \left\lceil \log_{3/2} n^{1/3} \right\rceil,
 \qquad
 \alpha := n^{1/(3\Delta)}.
\]
Then \(\alpha\in[5/4,3/2]\) for \(n^{1/3}\ge 2\), and
\(\alpha^\Delta=n^{1/3}\).

\paragraph{Phase 1: Centers and pivots.}
For \(i=0,1,\ldots,\Delta-1\), the algorithm samples two sets
\(S_i^c, S_i^p\subseteq V\) by including each vertex independently with probability
\(\alpha^i/n\). Thus \(\mathbb{E}[|S_i^c|] , \mathbb{E}[|S_i^p|]=\alpha^i\), and $\cup_i S_i^p$ and $ \cup S_i^c$ are independent.

For every sampled vertex \(s\in S_i^c \cup S_i^p\), the algorithm runs outgoing and incoming Dijkstra searches from \(s\) in the original graph \(G\). It stores the resulting distances by setting
\[
 \delta(s,x)\gets d(s\to x),
 \qquad
 \delta(x,s)\gets d(x\to s)
\]
for every \(x\in V\). It then uses the sampled vertices as possible proxies and updates, for every ordered pair \(v,u\in V\),
\[
 \delta(v,u)
 \gets
 \min_{s\in S_i^c \cup S_i^p}
 \bigl\{\delta(v,u),\,\delta(v,s)+\delta(s,u)\bigr\}.
\]

After the distances from and to \(S_i^c \cup S_i^p\) have been stored, the algorithm updates the pivot and center sets. For the pivot set, for every \(v\in V\), if there exists \(s\in S_i^p\) such that
\[
 \forall p\in P(v):\quad
 2d(v\to s)+d(s\to p)
 <
 2d(v\to p)+d(p\to s),
\]
then one such vertex \(s\) is added to \(P(v)\). This is the same update rule as in the basic pivot oracle.

The center set is updated similarly, but only using sampled outgoing neighbors. Namely, if there exists \(s\in N(v)\cap S_i^c\) such that
\[
 \forall c\in C(v):\quad
 2d(v\to s)+d(s\to c)
 <
 2d(v\to c)+d(c\to s),
\]
then one such vertex \(s\) is added to \(C(v)\). Centers are local: every center in \(C(v)\) is an outgoing neighbor of \(v\).

\paragraph{Phase 2: Sparsification.}
After the sampling phase, the algorithm sparsifies \(G\) in one shot. For every edge \((v,u)\in E\), it checks whether there exists a center \(c\in C(v)\) such that
\[
 2d(v\to c)+d(c\to u)
 \le
 2w(v,u)+d(u\to c).
\]
If so, the edge \((v,u)\) is deleted from \(H\).

The role of this step is twofold. First, it reduces the outdegree of the residual graph \(H\). Second, if an edge \((a,b)\) on a shortest path from \(v\) to \(u\) is deleted, then the path $v \to c \to u$, where \(c\in C(a)\) is the center responsible for the deletion, gives a good approximation for $v \to u$. Since distances through every sampled vertex were already aggregated into \(\delta\), this estimate is already stored.

\paragraph{Phase 3: Truncated searches on \(H\).} 
Finally, for every vertex \(v\in V\), the algorithm runs an outgoing Dijkstra search from \(v\) in the sparse graph \(H\), truncated by the pivot set \(P(v)\). When a vertex \(u\) is extracted from the priority queue, let \(\widehat d_H(v\to u)\) denote its current distance label. The algorithm scans the outgoing edges of \(u\) only if, for every \(p\in P(v)\), 
\[ 
2\widehat d_H(v\to u)+\delta(u,p) < 2\delta(v,p)+\delta(p,u). 
\] 
Let \(T_H(v)\) be the resulting truncated shortest-path tree. For every \(u\in T_H(v)\), the algorithm stores the distance found by the search: 
\( 
\delta(v,u)\gets \min\bigl\{\delta(v,u),d_{T_H(v)}(v\to u)\bigr\}. 
\)

The output estimates are the values \(\delta(v,u)\) for all ordered pairs. The construction is given in Algorithm~\ref{alg:3fast-apsp}.

\begin{algorithm}[H]
\SetAlgoNoEnd
\caption{Construction of a $3$-APSP}
\label{alg:3fast-apsp}
\DontPrintSemicolon

\textbf{Input:} Directed graph $G=(V,E,w)$ with $w\ge 0$ \\
\textbf{Output:} A table \(\delta\) of one-way estimates.

\medskip
Set \(
\Delta \gets \left\lceil \log_{3/2} n^{1/3} \right\rceil,
\qquad
\alpha \gets n^{1/(3\Delta)}.
\)

Set $H \gets G$, $P(v), C(v) \gets \emptyset$ for each $v \in V$.

\medskip
\textbf{Phase 1: Centers and pivots}

\For{$i \gets 0$ \KwTo $\Delta-1$}{
 Sample $S_i^c, S_i^p \subseteq V$ by including each vertex independently
 with probability $\alpha^i / n$.\;
 
 Run outgoing and incoming Dijkstra from each $s \in S_i^c,S_i^p$ in $G$ \;
 Set $\delta(s,v) \gets d(s \to v), \, \delta(v,s) \gets d(v \to s)$ for every $v \in V$ and $s \in S_i^c,S_i^p$\;

 Set $\delta(v, u) \gets \min_{s \in S_i^c \cup S_i^p}\{\delta(v, u),\, \delta(v,s) + \delta(s,u)\}$ for each $v,u \in V$\;

 \ForEach{$v \in V$}{
 \If{$\exists s \in S_i^p$ such that
 \[
 \forall p \in P(v):\;
 2d(v \to s) + d(s \to p)
 < 2d(v \to p) + d(p \to s)
 \]
 }{
 Add one such $s$ to $P(v)$.\;
 }

 \If{$\exists s \in N(v) \cap S_i^c$ such that
 \[
 \forall c \in C(v):\;
 2d(v \to s) + d(s \to c)
 < 2d(v \to c) + d(c \to s)
 \]
 }{
 Add one such $s$ to $C(v)$.\;
 }
 }

}

\medskip
\textbf{Phase 2: Sparsify $G$}

\ForEach{$v \in V$}{
 \ForEach{$u \in N(v)$}{
 \If{$\exists c \in C(v) $ such that
 \[
 2d(v \to c) + d(c \to u)
 \le 2w(v,u) + d(u \to c)
 \]
 }
 {
 Delete $(v,u)$ from $H$.
 }
}}

\medskip
\textbf{Phase 3: Truncated searches on $H$}

\ForEach{$v \in V$}{
    Run outgoing Dijkstra from $v$ on $H$; when a vertex $u$ with current
    distance label $\widehat d_H(v \to u)$ is extracted, scan its outgoing
    edges only if
    \[
        \forall p \in P(v):\;
        2\widehat d_H(v \to u) + \delta(u,p)
        <
        2\delta(v,p) + \delta(p,u)
    \]
    Let the resulting truncated tree be $T_H(v)$.\;
    Set
    $\delta(v,u) \gets
    \min\{\delta(v,u),d_{T_H(v)}(v \to u)\}$
    for all $u \in T_H(v)$\;
}

\Return $\delta$.

\end{algorithm}

\subsection{Analysis}

We start by analyzing the stretch.
\begin{lemma} \label{lem:3-fast-stretch}
Let $v,u\in V$. The output $\delta(v,u)$ provides
\[
d(v \to u) \le \delta(v,u) \le 2d(v \to u) + d(u \to v).
\]
\end{lemma}

\begin{proof}
 Let $\pi$ be the path from $v$ to $u$ in $G$. We consider three cases (see figure \ref{fig:fast-3-cases}).

\begin{itemize}
 \item[\textbf{Case 1:}] If $\pi \nsubseteq H$ (see figure \ref{fig:4-stretch} in Section \ref{sec:4-apsp}).

 This means that there was some edge $(a,b) \in \pi$ that got eliminated during the second phase, by some center $c \in C(a)$. So we know that
 \[
 2d(a \to c) + d(c \to b)
 \le 2w(a,b) + d(b \to c).
 \]
 But this naturally extends to $\pi$:
 \[
 2d(v \to c) + d(c \to u)
 \le 2d(v \to u) + d(u \to c)
 \]

 During the first phase, the algorithm stored
 \begin{align*}
 \delta(v,u) \leq d(v \to c) + d(c \to u) &\le 2d(v \to u) + d(u \to c) - d(v \to c) \\ &\le 2d(v \to u) + d(u \to v).
 \end{align*}

 \item[\textbf{Case 2:}] If $\pi \subseteq H$ but $ \pi \nsubseteq T_H(v)$ (see figure \ref{fig:3-stretch} in Appendix \ref{app:sec3}).

This means that there was a vertex $x \in \pi$ and a pivot $p \in P(v)$ such that
\[
2\delta(v,p)+\delta(p,x)
\le 2d_H(v\to x)+\delta(x,p).
\]
Since $\delta(\cdot,p)=d(\cdot,p)$ and $d_H(v\to x)=d_G(v\to x)$ as $\pi\subseteq H$, we have
\[
2d(v\to p)+d(p\to x)
\le 2d(v\to x)+d(x\to p),
\]
and this extends to $\pi$:
\[
2d(v\to p)+d(p\to u)
\le 2d(v\to u)+d(u\to p).
\]

 During the first phase, the algorithm stored
 \begin{align*}
 \delta(v,u) \leq d(v \to p) + d(p \to u) &\le 2d(v \to u) + d(u \to p) - d(v \to p) \\ &\le 2d(v \to u) + d(u \to v).
 \end{align*}

 \item[\textbf{Case 3:}] If $\pi \subseteq H \cap T_H(v)$, so $\delta(v,u) = d(v,u)$.
\end{itemize}
 
\end{proof}

Next, we analyze the running time. We first record two consequences of the
sparsification and truncation analyses. The first is analogous to
Lemma~\ref{lem:4do-edge-bound}.

\begin{lemma}\label{lem:3-fast-H-size1}
After the second phase ends, for every \(v\in V\),
\[
 \mathbb{E}\bigl[|N_H(v)|\bigr]=O(n^{2/3}).
\]
\end{lemma}

The second is analogous to Lemma~\ref{lem:3do-size-time}.

\begin{lemma}\label{lem:3-fast-H-size2}
For every \(v\in V\),
\[
 \mathbb{E}\bigl[|T_H(v)|\bigr]
 \le
 \mathbb{E}\bigl[|T_G(v)|\bigr]
 =
 O(n^{2/3}).
\]
\end{lemma}

We can now bound the running time.

\begin{lemma}
Algorithm~\ref{alg:3fast-apsp} takes \(\widetilde O(n^{7/3})\) time in expectation.
\end{lemma}

\begin{proof}
We analyze the three phases separately. In the first phase, the algorithm runs
incoming and outgoing Dijkstra searches from every sampled vertex. Since
\[
 \mathbb{E}\left[\sum_i |S_i^c|\right] , \mathbb{E}\left[\sum_i |S_i^p|\right]=O(n^{1/3}),
\]
this costs \(\widetilde O(mn^{1/3})\). In addition, for every sampled vertex \(s\), the algorithm updates all ordered pairs using the approximation \(d(v\to s)+d(s\to u)\). This costs
\[
 O\!\left(n^2\sum_i |S_i^c \cup S_i^p|\right)
 =
 O(n^{7/3})
\]
in expectation. Thus the first phase takes \(\widetilde O(n^{7/3})\) time.

The second phase scans the edges of \(G\) and checks the deletion condition
against the centers of the tail vertex. Since each center set has logarithmic
size in expectation, this phase takes \(\widetilde O(m)\le \widetilde O(n^2)\)
time.

Finally, in the third phase we run a truncated Dijkstra search from every
vertex in \(H\). Since \(T_H(v)\subseteq T_G(v)\), the number of edges scanned
by the search from \(v\) is at most
\[
 \sum_{u\in T_G(v)} |N_H(u)|.
\]
The set \(T_G(v)\) depends only on the pivot samples \(S_i^p\), whereas \(H\)
depends only on the independent center samples \(S_i^c\). Hence,
\[
\begin{aligned}
\mathbb{E}\!\left[\sum_{u\in T_G(v)} |N_H(u)|\right]
&=
\sum_{u\in V}
\Pr[u\in T_G(v)]\,
\mathbb{E}[|N_H(u)|] \\
&\le
O(n^{2/3})
\sum_{u\in V}\Pr[u\in T_G(v)] \\
&=
O(n^{2/3})\,
\mathbb{E}[|T_G(v)|]
=
O(n^{4/3}),
\end{aligned}
\]
where we used Lemmas~\ref{lem:3-fast-H-size1} and
\ref{lem:3-fast-H-size2}. Thus, one truncated search costs
\(\widetilde O(n^{4/3})\) in expectation, and all searches together cost
\(\widetilde O(n^{7/3})\).
Overall, the expected running time is \(\widetilde O(n^{7/3})\).
\end{proof}

\subsection{Generalizing the algorithm}

In the previous algorithm, we showed how to compute a stretch-\(3\) APSP approximation in \(\widetilde O(n^{7/3})\) time. This running time results from balancing the first phase, where we update the estimates for all ordered pairs through sampled vertices, with the third phase, where we run the final truncated searches on the sparsified graph.

Let us first restate this balance. Suppose that the algorithm samples
\(|\bigcup_i S_i^c|= |\bigcup_i S_i^p| =\widetilde O(n^\gamma)\) vertices, and uses them as centers and as pivots. Then, by Lemmas~\ref{lem:3-fast-H-size1}
and~\ref{lem:3-fast-H-size2}, for every \(v\in V\),
\[
 |N_H(v)|=\widetilde O(n^{1-\gamma})
 \qquad\text{and}\qquad
 |T_H(v)|=\widetilde O(n^{1-\gamma})
\]
in expectation. The first phase takes
\(
 \widetilde O(mn^\gamma+n^{2+\gamma})
\)
time: the term \(mn^\gamma\) is for the Dijkstra searches from and to the
sampled vertices, and the term \(n^{2+\gamma}\) is for updating all ordered
pairs through the sampled vertices. The third phase takes
\[
 \widetilde O\!\left(
 n\cdot n^{1-\gamma}\cdot n^{1-\gamma}
 \right)
 =
 \widetilde O(n^{3-2\gamma})
\]
time. Substituting \(\gamma=1/3\) gives the
\(\widetilde O(n^{7/3})\)-time algorithm.

When \(m=o(n^2)\), the Dijkstra part \(mn^\gamma\) in the first phase becomes smaller than the all-pairs update term \(n^{2+\gamma}\). This suggests using
different sampling rates for centers and pivots. The values
\[
 d(v\to s)+d(s\to u)
\]
need to be computed only in Case 1 of Lemma~\ref{lem:3-fast-stretch}, when a shortest \(v\to u\) path is destroyed by the sparsification and \(s\) is the center in charge of a deleted edge on that path. The pivots, on the other hand, are needed only in Case 2 of Lemma~\ref{lem:3-fast-stretch}, when the shortest path survives in \(H\) but the truncated search stops before reaching \(u\).

We therefore stop sampling centers earlier than we stop sampling pivots.
Let the algorithm sample pivots up to size $n^\gamma$, and let it stop
sampling centers at size $n^\beta$, where $\beta\le\gamma$.
We compute the all-pairs estimates through sampled centers only.
After the truncated searches, for every $v,u\in V$, we also update
\[
\delta(v,u)\gets\min\left\{\delta(v,u),
\min_{p\in P(v)}\{d(v\to p)+d(p\to u)\}\right\}.
\]
Since $|P(v)|=O(\log n)$, this takes $O(n^2\log n)$ time.

With these parameters, the first phase takes
\[
 \widetilde O(mn^\gamma+n^{2+\beta})
\]
time. The term \(mn^\gamma\) is the cost of the Dijkstra searches from and to all sampled pivots, while the term \(n^{2+\beta}\) is the cost of updating all ordered pairs through the sampled centers.

The sparsification bound of Lemma~\ref{lem:3-fast-H-size1} depends only on the centers. Hence, after stopping the center sampling at size \(n^\beta\), we get
\[
 |N_H(v)|=\widetilde O(n^{1-\beta})
\]
in expectation for every \(v\in V\). Since the original graph may already have average degree \(m/n\), the relevant residual degree bound is
\[
 \widetilde O\!\left(
 \min\left\{\frac mn,\;n^{1-\beta}\right\}
 \right).
\]
Similarly, the truncated-search bound of Lemma~\ref{lem:3-fast-H-size2} depends only on the pivots. Since pivot sampling continues up to size
\(n^\gamma\), we get
\[
 |T_H(v)|=\widetilde O(n^{1-\gamma})
\]
in expectation for every \(v\in V\). Therefore, the third phase takes
\[
 \widetilde O\!\left(
 n\cdot n^{1-\gamma}\cdot
 \min\left\{\frac mn,\;n^{1-\beta}\right\}
 \right)
\]
time.

Altogether, the running time is
\[
 \widetilde O\!\left(
 n^2
 +mn^\gamma
 +n^{2+\beta}
 +n\cdot n^{1-\gamma}\cdot
 \min\left\{\frac mn,\;n^{1-\beta}\right\}
 \right).
\]
This expression already contains the previous pivot-only bound: taking
\(\beta=0\) and \(\gamma=1/2\) gives
\[
 \widetilde O(n^2+m\sqrt n).
\]

It remains to optimize the regime where the sparsification improves over the
original average degree. Write \(m=n^\mu\). In this regime,
\[
 n^{1-\beta}\le \frac mn,
\]
so the third phase becomes
\[
 \widetilde O(n^{3-\gamma-\beta}).
\]
We balance the three main terms:
\[
 mn^\gamma
 =
 n^{2+\beta}
 =
 n^{3-\gamma-\beta}.
\]
Equivalently,
\[
 \mu+\gamma=2+\beta=3-\gamma-\beta.
\]
From \(2+\beta=3-\gamma-\beta\), we get
\[
 \gamma=1-2\beta.
\]
Combining this with \(\mu+\gamma=2+\beta\) gives
\[
 \beta=\frac{\mu-1}{3},
 \qquad
 \gamma=\frac{5-2\mu}{3}.
\]
The resulting running time is
\[
 \widetilde O(n^{2+\beta})
 =
 \widetilde O\!\left(n^{(5+\mu)/3}\right)
 =
 \widetilde O(m^{1/3}n^{5/3}).
\]

The condition that the sparsification improves over the original average
degree is
\[
 n^{1-\beta}\le \frac mn,
\]
or equivalently
\(
 \beta\ge 2-\mu.
\)
Substituting \(\beta=(\mu-1)/3\), this holds exactly when
\(
 \mu\ge \frac74,
\)
that is, when \(m\ge n^{7/4}\). For \(m\le n^{7/4}\), the choice
\(\beta=0\) and \(\gamma=1/2\) gives the better bound
\(\widetilde O(n^2+m\sqrt n)\).

We conclude that the generalized algorithm runs in
\[
 \widetilde O\!\left(
 n^2+\min\{m\sqrt n,\;m^{1/3}n^{5/3}\}
 \right)
\]
time. In particular, for dense graphs this is at most
\(
 \widetilde O(n^{7/3}).
\)

\section{Lower Bounds}
\label{sec:lb}

In this section we prove Theorems~\ref{th:bmm-lb}
and~\ref{th:oneway-bmm-lb}. Both reductions are from Boolean Matrix Multiplication (BMM). The first shows
that improving the roundtrip stretch from \(3\) to \(3-\varepsilon\) would
imply a BMM algorithm with essentially the same running time. The second shows
that improving the coefficient \(2\) in the one-way guarantee
\[
 2d(u\to v)+d(v\to u)
\]
would also imply such an improvement for BMM, even if the reverse distance is
allowed with an arbitrary constant coefficient.

Under the standard combinatorial BMM hypothesis, that for
every constant \(\gamma>0\), no combinatorial algorithm multiplies two
\(n\times n\) Boolean matrices in \(O(n^{3-\gamma})\) time, these reductions
rule out truly subcubic combinatorial algorithms for either task.

These reductions are also meaningful for algebraic algorithms. Boolean matrix
multiplication can be solved in \(n^{\omega}\) time, where
\(\omega<2.372\), and Zwick's algebraic techniques give a
\((1+\varepsilon)\)-approximation for directed APSP with polynomially bounded
weights in \(\widetilde O(n^\omega/\varepsilon)\) time~\cite{Zwick2002}.
Thus, an algorithm beating either of the two lower-bound thresholds in time
\(T(n)\) would yield a BMM algorithm in time \(T(O(n))+O(n^2)\). Achieving such a bound in \(o(n^\omega)\) time would improve the
current algebraic bound for BMM.

Our reductions are related in spirit to previous conditional lower bounds for
directed girth approximation. Dalirrooyfard and Vassilevska Williams~\cite{dalirrooyfard20}
proved, under the $k$-Cycle Hypothesis, that no $O(m^{2-\varepsilon})$-time algorithm can achieve a $(2-\delta)$-approximation for directed girth. Since the directed girth is
the minimum nonzero roundtrip distance, this also gives a corresponding
\((2-\delta)\)-barrier for roundtrip APSP. However, this barrier is most
relevant for sparse graphs: the hard instances have \(m=\Theta(n)\), and in
dense graphs the resulting bound is dominated by the cubic-time regime ($n^3$). Here we prove a different lower bound, based directly on BMM, showing
that for roundtrip APSP the stretch-\(3\) threshold is already inherent.

\subsection{Stretch below \(3\)}

\begin{proof}[Proof of Theorem~\ref{th:bmm-lb}]
Let \(A,B\in\{0,1\}^{n\times n}\). We construct a directed unweighted graph
\(G\) with \(O(n)\) vertices and \(O(n^2)\) edges.

The graph has four vertex sets
\[
 S=\{s_i:i\in[n]\},\qquad
 X=\{x_k:k\in[n]\},\qquad
 T=\{t_j:j\in[n]\},\qquad
 Y=\{y_k:k\in[n]\}.
\]
For every \(A_{ik}=1\), add the edges
\[
 s_i\to x_k
 \qquad\text{and}\qquad
 y_k\to s_i.
\]
For every \(B_{kj}=1\), add the edges
\[
 x_k\to t_j
 \qquad\text{and}\qquad
 t_j\to y_k.
\]

If \((AB)_{ij}=1\), then for some \(k\) we have
\(A_{ik}=B_{kj}=1\). Hence
\[
 d_G(s_i\to t_j) = 2
 \qquad\text{and}\qquad
 d_G(t_j\to s_i) = 2,
\]
and therefore
\[
 d_G(s_i\leftrightarrows t_j) = 4.
\]

Suppose now that \((AB)_{ij}=0\). Then there is no length-\(2\) path from
\(s_i\) to \(t_j\), nor from \(t_j\) to \(s_i\). Any longer path from \(S\)
to \(T\) must first go from \(S\) to \(T\), return through \(T\to Y\to S\),
and then go from \(S\) to \(T\) again; hence it has length at least \(6\).
Thus
\[
 d_G(s_i\to t_j)\ge 6
 \qquad\text{and}\qquad
 d_G(t_j\to s_i)\ge 6,
\]
where either distance may be infinite. Therefore
\[
 d_G(s_i\leftrightarrows t_j)\ge 12.
\]

We conclude that any \((3-\varepsilon)\)-approximation algorithm for
roundtrip APSP running in time \(T(n)\) yields an algorithm for Boolean
matrix multiplication running in time \(T(O(n))+O(n^2)\).

\end{proof}

\subsection{The one-way coefficient \(2\)}

We next prove that the coefficient \(2\) in the asymmetric one-way guarantee
cannot be improved, even if the reverse distance is allowed with an arbitrary
constant coefficient.

\begin{proof}[Proof of Theorem~\ref{th:oneway-bmm-lb}]
Let \(A,B\in\{0,1\}^{n\times n}\), and choose an integer
\[
 L > \max\left\{1,\frac{C}{2\varepsilon}\right\}.
\]
Since \(C\) and \(\varepsilon\) are constants, \(L\) is a constant.

We construct a directed unweighted graph \(G\) with \(O(n)\) vertices and
\(O(n^2)\) edges. The graph has source vertices
\[
 S=\{s_i:i\in[n]\},
\]
target vertices
\[
 T=\{t_j:j\in[n]\},
\]
and middle vertices \(x_k\), \(k\in[n]\).

For each \(k\), create vertices
\[
 a^1_k,\ldots,a^{L-1}_k
 \qquad\text{and}\qquad
 b^1_k,\ldots,b^{L-1}_k.
\]
Add the fixed chains
\[
 a^1_k\to a^2_k\to\cdots\to a^{L-1}_k\to x_k
\]
and
\[
 x_k\to b^1_k\to b^2_k\to\cdots\to b^{L-1}_k
\]
for every \(k\). For every \(A_{ik}=1\), add the edge
\[
 s_i\to a^1_k.
\]
For every \(B_{kj}=1\), add the edge
\[
 b^{L-1}_k\to t_j.
\]
Thus, every witness \(k\) with \(A_{ik}=B_{kj}=1\) gives a path
\[
 s_i \leadsto x_k \leadsto t_j
\]
of length \(2L\).

Finally, add all reverse edges
\[
 t_j\to s_i
 \qquad \forall i,j.
\]
Therefore
\[
 d_G(t_j\to s_i)=1
 \qquad \forall i,j.
\]

We claim that
\[
 (AB)_{ij}=1
 \quad\Longrightarrow\quad
 d_G(s_i\to t_j)\le 2L,
\]
whereas
\[
 (AB)_{ij}=0
 \quad\Longrightarrow\quad
 d_G(s_i\to t_j)\ge 4L.
\]
The first implication follows from the product path. For the second, suppose
\((AB)_{ij}=0\). Then there is no length-\(2L\) product path from \(s_i\) to
\(t_j\). Any longer path from \(s_i\) to \(t_j\) must first reach some target
\(t_{j'}\), use a reverse edge \(t_{j'}\to s_{i'}\), and then reach \(t_j\)
again. This costs at least
\[
 2L+1+2L > 4L.
\]
Thus \(d_G(s_i\to t_j)\ge 4L\), where the distance may be infinite.

Now run the assumed algorithm. If \((AB)_{ij}=1\), then
\[
 d_G(s_i\to t_j)\le 2L
 \qquad\text{and}\qquad
 d_G(t_j\to s_i)=1.
\]
Hence
\[
 \dist_{\mathcal D}(s_i,t_j)
 \le
 (2-\varepsilon)\cdot 2L + C
 =
 4L-2\varepsilon L+C
 <
 4L,
\]
by the choice of \(L\).

If \((AB)_{ij}=0\), then \(d_G(s_i\to t_j)\ge 4L\), and by
non-underestimation,
\[
 \dist_{\mathcal D}(s_i,t_j)\ge 4L.
\]
Therefore
\[
 (AB)_{ij}=1
 \quad\Longleftrightarrow\quad
 \dist_{\mathcal D}(s_i,t_j)<4L.
\]
Reading all \(n^2\) pairs \((s_i,t_j)\) gives the Boolean product \(AB\) in
time \(T(O(n))+O(n^2)\).
\end{proof}

\bibliographystyle{unsrt}
\bibliography{mybib}

@book{cormen01introduction,
  author = {Cormen, Thomas H. and Leiserson, Charles E. and Rivest, Ronald L. and Stein, Clifford},
  title = {{Introduction to Algorithms}},
  year = {2001},
  edition = {2nd},
  publisher = {MIT Press},
  isbn = {0262032937}
}

@article{Williams18,
  author = {Williams, R. Ryan},
  title = {{Faster All-Pairs Shortest Paths via Circuit Complexity}},
  journal = {SIAM Journal on Computing},
  volume = {47},
  number = {5},
  pages = {1965-1985},
  year = {2018},
  doi = {10.1137/15M1024524}
}

@article{Virginia18,
  author = {Vassilevska Williams, Virginia},
  title = {{Some Open Problems in Fine-Grained Complexity}},
  journal = {SIGACT News},
  volume = {49},
  number = {4},
  pages = {29--35},
  year = {2018},
  publisher = {Association for Computing Machinery},
  doi = {10.1145/3300150.3300158},
  address = {New York, NY, USA}
}

@article{AGM97,
  author = {Noga Alon and Zvi Galil and Oded Margalit},
  title = {{On the Exponent of the All Pairs Shortest Path Problem}},
  journal = {Journal of Computer and System Sciences},
  volume = {54},
  number = {2},
  pages = {255-262},
  year = {1997},
  doi = {10.1006/jcss.1997.1388}
}

@inproceedings{Alman25,
  author = {Alman, Josh and Duan, Ran and Vassilevska Williams, Virginia and Xu, Yinzhan and Xu, Zixuan and Zhou, Renfei},
  title = {{More Asymmetry Yields Faster Matrix Multiplication}},
  booktitle = {Proceedings of the 2025 Annual ACM-SIAM Symposium on Discrete Algorithms (SODA)},
  pages = {2005-2039},
  year = {2025},
  doi = {10.1137/1.9781611978322.63}
}

@article{Zwick2002,
  author = {Zwick, Uri},
  title = {{All pairs shortest paths using bridging sets and rectangular matrix multiplication}},
  journal = {J. ACM},
  volume = {49},
  number = {3},
  pages = {289--317},
  year = {2002},
  publisher = {Association for Computing Machinery},
  doi = {10.1145/567112.567114},
  address = {New York, NY, USA}
}

@article{Dor00,
  author = {Dor, Dorit and Halperin, Shay and Zwick, Uri},
  title = {{All-Pairs Almost Shortest Paths}},
  journal = {SIAM Journal on Computing},
  volume = {29},
  number = {5},
  pages = {1740-1759},
  year = {2000},
  doi = {10.1137/S0097539797327908}
}

@article{Baswana10,
  author = {Baswana, Surender and Kavitha, Telikepalli},
  title = {{Faster Algorithms for All-pairs Approximate Shortest Paths in Undirected Graphs}},
  journal = {SIAM Journal on Computing},
  volume = {39},
  number = {7},
  pages = {2865-2896},
  year = {2010},
  doi = {10.1137/080737174}
}

@inproceedings{Gupta25,
  author = {Manoj Gupta},
  title = {{Improved 2-Approximate Shortest Paths for close vertex pairs}},
  booktitle = {Proceedings of the 66th Annual IEEE Symposium on Foundations of Computer Science (FOCS 2025)},
  pages = {1248--1261},
  year = {2025},
  doi = {10.1109/FOCS63196.2025.00065}
}

@inproceedings{Jin26,
  author = {Jin, Ce and Kirkpatrick, Yael and Stawarz, Micha{\l} and Vassilevska Williams, Virginia},
  title = {{Improved Additive Approximation Algorithms for APSP}},
  booktitle = {Proceedings of the 2026 Annual ACM-SIAM Symposium on Discrete Algorithms (SODA)},
  pages = {3639-3651},
  year = {2026},
  doi = {10.1137/1.9781611978971.133}
}

@article{RS11,
  author = {Roditty, Liam and Shapira, Asaf},
  title = {{All-pairs shortest paths with a sublinear additive error}},
  journal = {ACM Trans. Algorithms},
  volume = {7},
  number = {4},
  year = {2011},
  publisher = {Association for Computing Machinery},
  doi = {10.1145/2000807.2000813},
  note = {Article 45},
  address = {New York, NY, USA}
}

@inproceedings{CowenWagner00,
  author = {Cowen, Lenore J. and Wagner, Christopher G.},
  title = {{Compact roundtrip routing in directed networks (extended abstract)}},
  booktitle = {Proceedings of the Nineteenth Annual ACM Symposium on Principles of Distributed Computing},
  series = {PODC '00},
  pages = {51--59},
  year = {2000},
  publisher = {Association for Computing Machinery},
  isbn = {1581131836},
  doi = {10.1145/343477.343516},
  address = {New York, NY, USA}
}

@article{RTZ08,
  author = {Roditty, Liam and Thorup, Mikkel and Zwick, Uri},
  title = {{Roundtrip spanners and roundtrip routing in directed graphs}},
  journal = {ACM Trans. Algorithms},
  volume = {4},
  number = {3},
  year = {2008},
  publisher = {Association for Computing Machinery},
  doi = {10.1145/1367064.1367069},
  note = {Article 29},
  address = {New York, NY, USA}
}

@inproceedings{HarbuzovaJinVWXu24,
  author = {Harbuzova, Alina and Jin, Ce and Vassilevska Williams, Virginia and Xu, Zixuan},
  title = {{Improved Roundtrip Spanners, Emulators, and Directed Girth Approximation}},
  booktitle = {Proceedings of the 2024 Annual ACM-SIAM Symposium on Discrete Algorithms (SODA)},
  pages = {4641-4669},
  year = {2024},
  doi = {10.1137/1.9781611977912.166}
}

@inproceedings{dalirrooyfard20,
  author = {Mina Dalirrooyfard and Virginia Vassilevska Williams},
  title = {{Conditionally optimal approximation algorithms for the girth of a directed graph}},
  booktitle = {47th International Colloquium on Automata, Languages, and Programming (ICALP 2020)},
  series = {Leibniz International Proceedings in Informatics (LIPIcs)},
  volume = {168},
  pages = {35:1--35:20},
  year = {2020},
  doi = {10.4230/LIPIcs.ICALP.2020.35}
}

@inproceedings{CL21,
  author = {Chechik, Shiri and Lifshitz, Gur},
  title = {{Optimal girth approximation for dense directed graphs}},
  booktitle = {Proceedings of the Thirty-Second Annual ACM-SIAM Symposium on Discrete Algorithms},
  series = {SODA '21},
  pages = {290--300},
  year = {2021},
  publisher = {Society for Industrial and Applied Mathematics},
  isbn = {9781611976465},
  address = {USA}
}

@article{plusTwo,
  author = {Aingworth, D. and Chekuri, C. and Indyk, P. and Motwani, R.},
  title = {{Fast Estimation of Diameter and Shortest Paths (Without Matrix Multiplication)}},
  journal = {SIAM Journal on Computing},
  volume = {28},
  number = {4},
  pages = {1167-1181},
  year = {1999},
  doi = {10.1137/S0097539796303421}
}

@inproceedings{roundtrip_spanners18,
  author = {Pachocki, Jakub and Roditty, Liam and Sidford, Aaron and Tov, Roei and Vassilevska Williams, Virginia},
  title = {{Approximating Cycles in Directed Graphs: Fast Algorithms for Girth and Roundtrip Spanners}},
  booktitle = {Proceedings of the 2018 Annual ACM-SIAM Symposium on Discrete Algorithms (SODA)},
  pages = {1374-1392},
  year = {2018},
  doi = {10.1137/1.9781611975031.91}
}

@inproceedings{roundtrip_spanners19,
  author = {Shiri Chechik and
               Yang P. Liu and
               Omer Rotem and
               Aaron Sidford},
  title = {{Constant girth approximation for directed graphs in subquadratic time}},
  booktitle = {Proceedings of the 52nd Annual ACM SIGACT Symposium on Theory of Computing (STOC 2020)},
  pages = {1010--1023},
  year = {2020},
  doi = {10.1145/3357713.3384330}
}

@inproceedings{roundtrip_spanners20b,
  author = {Ruoxu Cen and
               Ran Duan and
               Yong Gu},
  title = {{Roundtrip Spanners with (2k-1) Stretch}},
  booktitle = {47th International Colloquium on Automata, Languages, and Programming (ICALP 2020)},
  series = {Leibniz International Proceedings in Informatics (LIPIcs)},
  volume = {168},
  pages = {24:1--24:11},
  year = {2020},
  doi = {10.4230/LIPIcs.ICALP.2020.24}
}

@inproceedings{Fischer1971BooleanMM,
  author = {Michael J. Fischer and Albert R. Meyer},
  title = {{Boolean Matrix Multiplication and Transitive Closure}},
  booktitle = {Proceedings of the 12th Annual Symposium on Switching and Automata Theory (SWAT 1971)},
  pages = {129--131},
  year = {1971},
  doi = {10.1109/SWAT.1971.4}
}

@article{VW10,
  author = {Vassilevska Williams, Virginia and Williams, R. Ryan},
  title = {{Subcubic Equivalences Between Path, Matrix, and Triangle Problems}},
  journal = {J. ACM},
  volume = {65},
  number = {5},
  year = {2018},
  publisher = {Association for Computing Machinery},
  doi = {10.1145/3186893},
  note = {Article 27},
  address = {New York, NY, USA}
}

@article{CZ01,
  author = {Cohen, Edith and Zwick, Uri},
  title = {{All-Pairs Small-Stretch Paths}},
  journal = {Journal of Algorithms},
  volume = {38},
  number = {2},
  pages = {335--353},
  year = {2001},
  doi = {10.1006/jagm.2000.1117}
}

@inproceedings{Rod23,
  author = {Roditty, Liam},
  title = {{New Algorithms for All Pairs Approximate Shortest Paths}},
  booktitle = {Proceedings of the 55th Annual ACM Symposium on Theory of Computing (STOC)},
  pages = {309--320},
  year = {2023},
  doi = {10.1145/3564246.3585197}
}

@inproceedings{DFK24,
  author = {Dory, Michal and Forster, Sebastian and Kirkpatrick, Yael and Nazari, Yasamin and Vassilevska Williams, Virginia and de Vos, Tijn},
  title = {{Fast 2-Approximate All-Pairs Shortest Paths}},
  booktitle = {Proceedings of the 2024 Annual ACM-SIAM Symposium on Discrete Algorithms (SODA)},
  pages = {4728--4757},
  year = {2024},
  doi = {10.1137/1.9781611977912.169}
}

@inproceedings{SY24,
  author = {Saha, Barna and Ye, Christopher},
  title = {{Faster Approximate All Pairs Shortest Paths}},
  booktitle = {Proceedings of the 2024 Annual ACM-SIAM Symposium on Discrete Algorithms (SODA)},
  pages = {4758--4827},
  year = {2024},
  doi = {10.1137/1.9781611977912.170}
}

@article{Kavitha12,
  author = {Kavitha, Telikepalli},
  title = {{Faster Algorithms for All-Pairs Small Stretch Distances in Weighted Graphs}},
  journal = {Algorithmica},
  volume = {63},
  number = {1--2},
  pages = {224--245},
  year = {2012},
  doi = {10.1007/s00453-011-9529-y}
}

@inproceedings{AR21,
  author = {Akav, Maor and Roditty, Liam},
  title = {{A Unified Approach for All Pairs Approximate Shortest Paths in Weighted Undirected Graphs}},
  booktitle = {29th Annual European Symposium on Algorithms (ESA 2021)},
  series = {Leibniz International Proceedings in Informatics (LIPIcs)},
  volume = {204},
  pages = {4:1--4:18},
  year = {2021},
  doi = {10.4230/LIPIcs.ESA.2021.4}
}

\appendix

\section{3-Roundtrip Distance Oracle: Analysis}
\label{app:sec3}

In this appendix, we provide the full analysis of Algorithm~\ref{alg:3do-construct}.
We begin by analyzing the stretch guarantee of the oracle.

\begin{lemma}
 For any $v,u \in V$, 
\[ 
d(v \to u) \le \dist_{\mathcal{D}}(v,u) \le 2\,d(v \to u) + d(u \to v).
\]
\end{lemma}

\begin{proof}
 Fix any query $(v,u)$, and let $Q$ be a shortest path from $v$ to $u$ in $G$. The oracle returns
 \[
\dist_{\mathcal{D}}(v,u)
 \;=\;
 \min_{p\in P(v)} \bigl\{ \delta(v,p) + \delta(p,u) \bigr\}.
\]
If all vertices of $Q$ are explored by the truncated Dijkstra search from $v$, then $Q \subseteq T(v)$ and therefore $\delta(v,u)=d(v \to u)$. So,
\[
\dist_{\mathcal{D}}(v,u)
 \;=\;
 \min_{p\in P(v)} \bigl\{ \delta(v,p) + \delta(p,u) \bigr\} = \delta(v,v) + \delta(v,u) = d(v \to u),
\]
as $v \in P(v)$.
Otherwise, let $x$ be the first vertex on $Q$ (closest to $v$) that is not relaxed during the truncated search. By definition of the truncation predicate, there exists a pivot $p \in P(v)$ such that
\begin{equation}\label{eq:fail-x}
2d(v \to p)+d(p \to x)
\;\le\;
2d(v \to x)+d(x \to p).
\end{equation}

Add $d(x\to u)$ to both sides of~\eqref{eq:fail-x}:
\[
2d(v \to p)+d(p \to x)+d(x\to u)
\;\le\;
2d(v \to x)+d(x \to p)+d(x\to u).
\]
By the triangle inequality,
\[
\text{(A)}\qquad d(p\to u)\le d(p\to x)+d(x\to u),
\qquad\qquad
\text{(B)}\qquad d(x\to p)\le d(x\to u)+d(u\to p).
\]
Hence
\begin{align*}
2d(v \to p) + d(p \to u)
&\le 2d(v \to p) + \bigl(d(p\to x)+d(x\to u)\bigr) && \text{by (A)}\\
&\le 2d(v \to x) + \bigl(d(x\to p)+d(x\to u)\bigr) && \text{by \eqref{eq:fail-x}}\\
&\le 2d(v \to x) + \bigl(d(x\to u)+d(u\to p)+d(x\to u)\bigr) && \text{by (B)}\\
&= 2d(v \to x) + 2d(x\to u) + d(u\to p).
\end{align*}
Since $x$ lies on a shortest $v\to u$ path, $d(v\to u)=d(v\to x)+d(x\to u)$, and therefore
\begin{equation}
 \label{eq:3-stretch-x-to-u}
 2d(v \to p)+d(p \to u)
\;\le\;
2d(v\to u)+d(u\to p),
\end{equation}
and the inequality naturally extends from $x$ to $u$, see Figure \ref{fig:3-stretch}.

\begin{figure}[h]
 \centering
 \begin{tikzpicture}

 \node[circle,draw,inner sep=2.5pt,label=below left:$v$] (v) at (0,-2) {};
\node[circle,draw,inner sep=2.5pt,label=below:$x$] (x) at (1.25,-0.5) {};
\node[circle,draw,inner sep=2.5pt,label=left:$p$] (p) at (-1.25,-0.5) {};

\node[circle,draw,inner sep=2.5pt,label=left:$u$] (u) at (2.5,1) {};

\draw[directed] (v) to (p);
\draw[line width=4pt, green, opacity=0.3] (v) to (p);

\draw[directed, purple] (v) to (x);

\draw[directed, bend left=30] (p) to (x);
\draw[bend left=30,line width=4pt, green, opacity=0.3] (p) to (x);

\draw[directed, bend left=30] (x) to (p);

\draw[directed, purple] (x) to (u);
\draw[line width=4pt, green, opacity=0.3] (x) to (u);
\draw[directed, bend left=50] (u) to (v);
 
 \end{tikzpicture}
 \caption{An illustration of the bypass intuition underlying Inequality~\ref{eq:3-stretch-x-to-u}. 
When the truncation condition fails at vertex $x$ (Inequality~\ref{eq:fail-x}), the path from $v$ to $u$ (purple) is conceptually rerouted via the pivot $p$, following a path of the form $v\!\to\!p\!\to\!x\!\to\!u$ (green).}
 \label{fig:3-stretch}
\end{figure}
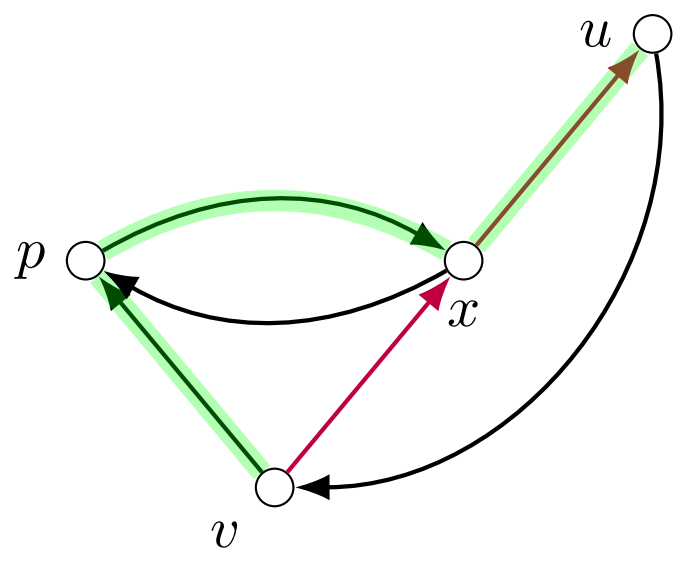

Subtracting $d(v \to p)$ from both sides yields
\[
d(v \to p) + d(p\to u)
\;\le\;
2d(v \to u) + d(u \to p) - d(v \to p)
\;\le\;
2d(v \to u) + d(u \to v).
\]
Therefore,
\[
\dist_{\mathcal{D}}(v,u)\;\le\;2d(v \to u) + d(u \to v).
\]

\end{proof}

Next, we bound the expected size stored in the oracle. To do so, we first bound the expected size of $T(v)$ for every $v$.

\begin{definition}[Eligible pivots]
Fix $v\in V$. For each iteration $i\ge 0$, let $P_i(v)$ be the pivot set of $v$ at the beginning of iteration $i$. Define the set of eligible pivots for $v$ at iteration $i$ by
\[
B_i(v)=\Bigl\{u\in V:\ \forall p\in P_i(v),\ 
2d(v\to u)+d(u\to p)<2d(v\to p)+d(p\to u)\Bigr\}.
\]
In other words, $B_i(v)$ consists of the vertices that can still be selected as future pivots for $v$ after the first $i$ rounds.
\end{definition}

At iteration $i$, a new pivot is added to $P(v)$ if and only if $S_i\cap B_i(v)\neq\emptyset$, since a sampled vertex is eligible to be chosen as a pivot for $v$ exactly when it belongs to $B_i(v)$.

\begin{claim}\label{clm:Bi}
Fix $v\in V$. Then for all $i$,
\[
\mathbb{E}[|B_i(v)|]\ \le\ \frac{\alpha}{2-\alpha}\cdot \frac{n}{\alpha^i}\ \le\ 3\cdot \frac{n}{\alpha^i}.
\]
\end{claim}

Before proving the claim, observe that the truncated search tree $T(v)$ is always a subset of $B_\Delta(v)$, since the algorithm relaxes a vertex only if it satisfies the truncation predicate with respect to all pivots in $P(v)$ after all $\Delta$ iterations. Therefore, by setting $i=\Delta$ and using that $\alpha^\Delta=\sqrt{n}$, Claim~\ref{clm:Bi} implies
\[
\mathbb{E}[|T(v)|] \;\le\; \mathbb{E}[|B_\Delta(v)|] \;=\; O(\sqrt{n})
\qquad\text{for every } v\in V.
\]

\begin{proof}
Fix $i$ and $v$, and let $p_i:=\alpha^i/n$ be the sampling probability of $S_i$ in iteration $i$. Assume we are at iteration $i$ of the algorithm, so that $B_i(v)$ is fixed, and we wish to bound \(\mathbb{E}[|B_{i+1}(v)|\mid B_i(v)]\).

Consider the complete directed graph on the vertex set $B_i(v)$ defined as follows. For every unordered pair
$x,y\in B_i(v)$, orient the edge from $x$ to $y$ if
\[
2d(v\to y)+d(y\to x)\ \ge\ 2d(v\to x)+d(x\to y),
\]
and otherwise orient it from $y$ to $x$ (breaking ties arbitrarily). Thus, this orientation defines a tournament on $B_i(v)$.

Observe that if a vertex $s\in B_i(v)$ is added to the pivot set of $v$, then every vertex $y\in B_i(v)$
such that the edge is oriented from $s$ to $y$ is removed from the set of eligible pivots in the next iteration.
Indeed, once $s$ is added to the pivot set, such a vertex $y$ no longer satisfies the defining inequality for membership in $B_{i+1}(v)$. Conversely, every vertex $y\in B_i(v)\setminus\{s\}$ such that the edge is oriented from $y$ to $s$ remains in $B_{i+1}(v)$, since it still satisfies the defining inequalities for all pivots in $P_i(v)$, and also for the new pivot $s$. Therefore, after adding $s$, the set $B_{i+1}(v)$ is exactly the set of in-neighbors of $s$ in this tournament.

In particular, if one chooses a vertex uniformly at random from $B_i(v)$, then the expected size of the resulting set $B_{i+1}(v)$ is
\[
\frac{1}{|B_i(v)|}\sum_{x\in B_i(v)} \deg^-(x)
\;=\;
\frac{|B_i(v)|-1}{2},
\]
since a tournament on $|B_i(v)|$ vertices has exactly $\binom{|B_i(v)|}{2}$ directed edges.

Thus, conditioning on the event $S_i\cap B_i(v)\neq\emptyset$, the new pivot chosen for $v$ is distributed uniformly over $B_i(v)$ by symmetry: conditioned on this event, each vertex of $B_i(v)$ has the same probability to be selected, since the vertices are sampled independently and the algorithm chooses uniformly among the sampled eligible vertices. Hence,
\[
\mathbb{E}\bigl[\,|B_{i+1}(v)| \,\bigm|\, B_i(v),\, S_i\cap B_i(v)\neq\emptyset \bigr]
\;=\; \frac{|B_i(v)|-1}{2}
\;\le\; \frac{|B_i(v)|}{2},
\]
where the expectation is on the choice of $S_i$.

On the other hand, if $S_i\cap B_i(v)=\emptyset$, then no new pivot is added for $v$ in iteration $i$, and therefore
\[
B_{i+1}(v)=B_i(v).
\]
Since each vertex of $B_i(v)$ is sampled independently with probability $p_i$, we have
\[
\Pr\bigl[S_i\cap B_i(v)=\emptyset \,\bigm|\, B_i(v)\bigr]=(1-p_i)^{|B_i(v)|}.
\]
Combining the two cases by conditioning on the event $S_i\cap B_i(v)=\emptyset$, we obtain
\[
\mathbb{E}\bigl[\,|B_{i+1}(v)| \,\bigm|\, B_i(v)\bigr]
\;\le\;
|B_i(v)|\cdot (1-p_i)^{|B_i(v)|}
\;+\;
\frac{|B_i(v)|}{2}\cdot \bigl(1-(1-p_i)^{|B_i(v)|}\bigr),
\]
and hence
\[
\mathbb{E}\bigl[\,|B_{i+1}(v)| \,\bigm|\, B_i(v)\bigr]
\;\le\;
\frac{|B_i(v)|}{2}\Bigl(1+(1-p_i)^{|B_i(v)|}\Bigr).
\]

We now convert this inequality into a recurrence. Multiplying both sides by
\[
p_{i+1}=\alpha p_i
\]
and using the bound
\[
(1-p_i)^x\le e^{-p_i x}\qquad\text{for all }x\ge 0,
\]
gives
\[
\mathbb{E}\bigl[\,p_{i+1}|B_{i+1}(v)| \,\bigm|\, B_i(v)\bigr]
\;\le\;
\frac{\alpha}{2}\, p_i |B_i(v)|\Bigl(1+e^{-p_i|B_i(v)|}\Bigr).
\]
Using also the standard inequality
\[
xe^{-x}\le 1\qquad\text{for all }x\ge 0,
\]
we get
\[
\mathbb{E}\bigl[\,p_{i+1}|B_{i+1}(v)| \,\bigm|\, B_i(v)\bigr]
\;\le\;
\frac{\alpha}{2}\Bigl(p_i|B_i(v)|+1\Bigr).
\]

Let
\[
X_i:=p_i\,\mathbb{E}[|B_i(v)|].
\]
Taking expectations yields
\[
X_{i+1}\le \frac{\alpha}{2}(X_i+1).
\]
Note that $X_0 = 1$, so implies by induction that
\[
X_i\le \frac{\alpha}{2-\alpha}
\qquad\text{for all }i\ge 0.
\]
Therefore,
\[
\mathbb{E}[|B_i(v)|]
=
\frac{X_i}{p_i}
\le
\frac{\alpha}{2-\alpha}\cdot \frac{n}{\alpha^i},
\]
as claimed. Since $\alpha\le 3/2$, we also have
\[
\frac{\alpha}{2-\alpha}\le 3.
\]
\end{proof}

\begin{lemma}
\label{lem:3do-size-time}
The size of the oracle is $O(n^{3/2})$ in expectation, and the preprocessing time is $O(m \sqrt{n}\log n)$ in expectation.
\end{lemma}

\begin{proof}
We bound the space usage and preprocessing time of each phase separately.

\paragraph{Phase~1.}
In iteration $i<\Delta$, for every sampled witness $s\in S_i$ the algorithm stores the distances
between $s$ and all vertices $v\in V$.
Since $\mathbb{E}[|S_i|]=\alpha^i$ and $\alpha^\Delta=\sqrt{n}$, the expected total amount of
information stored in Phase~1 is
\[
\mathbb{E}\!\left[\sum_{i=0}^{\Delta-1} |V|\cdot |S_i|\right]
\;=\;
n \sum_{i=0}^{\Delta-1} \mathbb{E}[|S_i|]
\;=\;
O(n^{3/2}).
\]
The expected running time of this phase is $O(m\sqrt{n})$, dominated by the incoming and outgoing
Dijkstra searches performed from all sampled witnesses
$s\in\bigcup_{i<\Delta} S_i$.

\paragraph{Phase~2.}
For each vertex $v\in V$, the truncated Dijkstra search from $v$ explores a set
$T(v)\subseteq V$ of vertices satisfying the truncation predicate.
By Claim~\ref{clm:Bi}, we have $\mathbb{E}[|T(v)|]=O(\sqrt{n})$ for every $v$.
Since the oracle stores the distances $\delta(v,u)$ for all $u\in T(v)$, the total expected amount
of information stored in Phase~2 is
\[
\sum_{v\in V} \mathbb{E}[|T(v)|] = O(n^{3/2}).
\]

This phase runs in expected time $O(m\sqrt{n}\log n)$.
For each vertex $v$, the truncated Dijkstra search explores $|T(v)|=O(\sqrt{n})$ vertices in expectation.
Assuming the maximum out-degree is $O(m/n)$, the cost of running Dijkstra restricted to $T(v)$ is
$O\!\left(\tfrac{m}{n}\cdot |T(v)| \log n\right)=O(m/\sqrt{n}\log n)$.
In addition, each relaxation requires checking the truncation predicate against all pivots in $P(v)$,
which takes $O(|P(v)|)=O(\log n)$ time.
Summing over all $v\in V$ yields a total expected running time of $O(m\sqrt{n}\log n)$.

The assumption that the maximum degree is $O(m/n)$ is without loss of generality: as shown by Chechik et al.~\cite{roundtrip_spanners19}, any directed graph can be reduced to this case in $O(m)$ time.

\end{proof}

\section{A $5$-Roundtrip APSP Approximation in $\tilde{O}(n^2)$}
\label{app:5-apsp}

A natural question in the study of approximate APSP in directed graphs is what
roundtrip approximation factor can be achieved within $O(n^2)$ time, which is a
fundamental lower bound imposed by the output size of the problem. While the
algorithms in the previous sections achieve a small roundtrip stretch, their
running times exceed this quadratic barrier.

In this section, we show that by allowing a slightly larger approximation factor,
this barrier can be met up to polylogarithmic factors. Specifically, we prove
Theorem~\ref{thm:5-roundtrip-do} and Corollary~\ref{cor:5-roundtrip-apsp} by
constructing a $5$-roundtrip distance oracle with near-quadratic preprocessing
time, and showing how querying it on all vertex pairs yields a $5$-APSP
approximation in expected $\tilde{O}(n^2)$ time.

At a high level, the distance oracle construction combines the
center-based sparsification of Section~\ref{sec:4-apsp} with the
pivot-based truncated-search framework of Section~\ref{sec:3-apsp}.
The algorithm first applies the center-sampling process, progressively
removing edges while recording limited pivot information, until the
remaining graph becomes sufficiently sparse. Once the number of edges
drops to \(O(n^{3/2})\) in expectation, the algorithm applies the
\(3\)-roundtrip oracle construction of Section~\ref{sec:3-apsp} to the
sparsified graph.
The phase transition is chosen so that the second phase can be executed
within the quadratic time budget. Indeed, when applied to a graph with
\(O(n^{3/2})\) edges, the preprocessing time of the \(3\)-roundtrip oracle is
\(
 \widetilde{O}(m\sqrt{n})=\widetilde{O}(n^2).
\)

\subsection{Construction Overview}

The construction proceeds in two phases. 
The first phase follows the sparsification framework of the $4$-roundtrip APSP
algorithm from Section~\ref{sec:4-apsp}, but replaces the expensive aggregation
step by selecting a single pivot per vertex in each iteration.
The second phase applies the $3$-roundtrip distance oracle construction on the
resulting sparse graph.

\paragraph{Phase 1: Sparsification with nearest pivots.}
The algorithm maintains a sequence of graphs
$G_0, G_1, \dots$, obtained by iteratively deleting edges.
Initially, $G_0 := G$.

It performs
\[
\Delta := \left\lceil \log_{3/2} \sqrt{n} \right\rceil
\quad\text{iterations, where}\quad
\alpha := (\sqrt{n})^{1/\Delta}.
\]
Note that $\alpha \in [5/4,3/2]$ for $n \ge 4$, and $\alpha^\Delta = \sqrt{n}$.

For each vertex $v \in V$, the algorithm maintains a set of pivots $P(v)$,
initially empty. By the end of the first phase, $|P(v)| = O(\log n)$.

For iterations $i=0,1,\ldots,\Delta-1$, the following steps are executed.

\paragraph{Sampling pivots/centers.}
A set $S_i \subseteq V$ is sampled by including each vertex independently with
probability $\alpha^i / n$. Hence $\mathbb{E}[|S_i|] = \alpha^i$.

\paragraph{Computing distances and selecting pivots.}
Let $G_i$ denote the current graph.
For each sampled vertex $s \in S_i$, the algorithm runs both outgoing and incoming Dijkstra searches rooted at $s$ in the current graph $G_i$. Then, the oracle stores these distances by setting $\delta(s,v) \gets \min\{ \delta(s,v), d_i(s \to v) \}$ and $ \delta(v,s) \gets \min \{ \delta(v,s), d_i(v \to s) \}$ for all $v\in V$ and $s\in S_i$.

For each vertex $v \in V$, the algorithm selects a pivot
\[
p_i(v) \;:=\; \arg\min_{s \in S_i} d_i(v \lra s),
\]
and adds $p_i(v)$ to $P(v)$ (ties are broken arbitrarily).

This is the main departure from the $4$-roundtrip APSP algorithm:
instead of aggregating distances for all triples $v \lra s \lra u$, the algorithm keeps
only the nearest sampled pivot for each vertex.
As a result, the running time of this step is reduced from
$\Theta(n^2 \cdot |S_i|)$ to $\Theta(n \cdot |S_i|)$ per iteration.
We show later that this weaker aggregation still yields a sufficiently good roundtrip approximation.

\paragraph{Edge pruning.}
Edge pruning is performed exactly as in the $4$-roundtrip APSP algorithm.
For each vertex $v \in V$, let $N_i(v)$ denote its outgoing neighbors in $G_i$.
If $N_i(v) \cap S_i \neq \emptyset$, a center
$s_i(v) \in N_i(v) \cap S_i$ is chosen uniformly at random. Note that \(s_i(v)\) is the center used for pruning the outgoing edges of \(v\), whereas \(p_i(v)\) is the nearest pivot to \(v\); the two may differ.

The algorithm then inspects every edge $(v,u) \in E(G_i)$.
If $v$ has an assigned center $s_i(v)$ and
\[
 2\,d_i(v \to s_i(v)) + d_i(s_i(v) \to u)
 \;\leq\;
 2\,w(v,u) + d_i(u \to s_i(v)),
\]
the edge $(v,u)$ is deleted.
Let $G_{i+1}$ denote the resulting graph. Note that all the distances above are stored via $\delta$.

After $\Delta$ iterations, the first phase ends, and the graph $G_\Delta$ has $O(n^{3/2})$ edges in expectation.

\paragraph{Phase 2: Distance oracle on the sparse graph.}
In the second phase, the algorithm applies Algorithm~\ref{alg:3do-construct} to the
sparsified graph $G_\Delta$, as described in Section~\ref{sec:3-apsp}, and merges
the pivot sets produced by that algorithm with the pivot sets maintained in Phase~1.

\paragraph{Distance query.}
Similarly to Algorithm \ref{alg:3do-query}, to answer a query $\dist_{\mathcal{D}}(v,u)$, the oracle returns the minimum among the distances provided by the truncated tree of $v$ and by routing through the pivots of $v$:
\[
\dist_{\mathcal{D}}(v,u)
 \;=\;
 \min_{p\in P(v)}\Bigl\{ \delta(v,p)+\delta(p,u)
 \Bigr\}.
\]
We summarize the construction in
Algorithm ~\ref{alg:5-do}.

\begin{algorithm}[H]
\caption{Construction of a $5$-roundtrip distance oracle}
\label{alg:5-do}
\DontPrintSemicolon
\SetAlgoNoEnd

\KwIn{Directed graph $G=(V,E,w)$ with non-negative weights.}
\KwOut{Distance oracle $\mathcal{D}$}

$G_0 \gets G$;\quad $P(v) \gets \emptyset$ for each $v \in V$; \;

Set $\Delta \gets \left\lceil \log_{3/2} \sqrt{n} \right\rceil$ and $\alpha \gets (\sqrt{n})^{1/\Delta}$; \;

\For{$i \gets 0$ \KwTo $\Delta-1$}{
 Sample $S_i \subseteq V$ by including each $v \in V$ independently with probability $\alpha^i / n$; \;

 Run outgoing and incoming Dijkstra from each $s \in S_i$ in $G_i$ \;
 Set $\delta(s,v) \gets \min\{ \delta(s,v), d_i(s \to v) \}, \, \delta(v,s) \gets \min \{ \delta(v,s), d_i(v \to s) \}$ for every $v \in V$ and $s \in S_i$\;

 \ForEach{$v \in V$}{
 Set
 \(
 p_i(v) \gets \arg\min_{s \in S_i} \{ d_i(v \lra s )\};
 \)
 
 Add $p_i(v)$ to $P(v)$;
 }

 $G_{i+1} \gets G_i$; \;

 \ForEach{$v \in V$ such that $N_i(v) \cap S_i \neq \emptyset$}{
 pick a random $s_i(v) \in N_i(v) \cap S_i$; \;
 \ForEach{$u \in N_i(v)$}{
 \If{$2 d_i(v \to s_i(v)) + d_i(s_i(v) \to u) \le 2w(v,u) + d_i(u \to s_i(v))$}{
 delete edge $(v,u)$ from $G_{i+1}$; \;
 }
 }
 }
}

Run Algorithm \ref{alg:3do-construct} on $G_\Delta$ and merge the pivot sets for all $v \in V$.

\end{algorithm}

\subsection{Analysis}

We begin the proof of Theorem~\ref{thm:5-roundtrip-do} by analyzing the stretch of the distance oracle.

\begin{lemma}
For every $v,u \in V$, the distance query $\dist_{\mathcal{D}}(v,u)$ returns an estimate satisfying
\[
d(v \to u)
\;\leq\;
\dist_{\mathcal{D}}(v,u)
\;\leq\;
3d(v \to u) + 2d(u \to v).
\]
\end{lemma}

\begin{proof}
Fix vertices $v,u \in V$, and let $C$ denote a shortest roundtrip (cycle) between $v$ and $u$ in $G$.

If $C \subseteq G_\Delta$, then Algorithm \ref{alg:3do-construct} is applied to a graph containing the entire shortest roundtrip. As shown previously, in this case the oracle returns an estimate satisfying
\[
d(v \to u)
\;\leq\;
\dist_{\mathcal{D}}(v,u)
\;\leq\;
2d(v \to u) + d(u \to v),
\]
which is stronger than the claimed bound.

We therefore assume that $C \nsubseteq G_\Delta$.
Let $i$ be the last iteration such that $C \subseteq G_i$.
Then there exists an edge $e=(a,b)$ on $C$ with
\[
e \in G_i \setminus G_{i+1}.
\]

The edge $(a,b)$ was removed in iteration $i$ because there exists a vertex
$s_i(a):=s \in N_i(a) \cap S_i$ such that
\begin{equation*}
2d_i(a \to s) + d_i(s \to b)
\;\leq\;
2w(a,b) + d_i(b \to s).
\end{equation*}

Exactly as in the proof of Lemma ~\ref{lem:roundtrip-stretch-general}, one can show that~\eqref{eq:4-approx} holds (see Figure~\ref{fig:4-stretch}):
\[
d_i(v \lra s),\; d_i(s \lra u) \;\le\; 2\,d_i(v \lra u).
\]

By definition, the pivot $p_i(v)$ is chosen to be at least as close to $v$ as $s$, and hence
\[
d_i(v \lra p_i(v)) \le d_i(v \lra s).
\]

The oracle therefore routes from $v$ to $u$ via the pivot $p_i(v)$, yielding
\begin{align*}
\dist_{\mathcal{D}}(v,u)
&\le d_i(v \to p_i(v)) + d_i(p_i(v) \to u) \\
&\le d_i(v \lra p_i(v)) + d_i(v \to u) \\
&\le 2\,d_i(v \lra u) + d_i(v \to u),
\end{align*}
as claimed.
\end{proof}

Next, we analyze the size of the oracle and the preprocessing time.

\begin{lemma}
The oracle stores $O(n^{3/2})$ expected data and is constructed in expected time
$\tilde{O}(n^2)$.
\end{lemma}

\begin{proof}
We analyze the two phases separately. For the second phase, the bounds follow directly from
Lemma~\ref{lem:3do-size-time}, applied to the graph $G_\Delta$.
By Lemma~\ref{lem:4do-edge-bound}, we have $\mathbb{E}[|E(G_\Delta)|]=O(n^{3/2})$,
and since the same edge-pruning procedure is used here, the analysis carries over
without modification.

It therefore remains to bound the space usage and running time of the first phase. In this phase, the oracle stores distances computed by running Dijkstra from all sampled
centers in $\bigcup_{i<\Delta} S_i$.
Thus, the space usage is proportional to the total number of sampled centers, and the
running time is dominated by the corresponding Dijkstra executions.

Since $\mathbb{E}[|S_i|]=\alpha^i$ and $\sum_{i<\Delta}\alpha^i = O(\sqrt{n})$, we have
\[
\mathbb{E}\bigl[|\textstyle\bigcup_{i<\Delta} S_i|\bigr]=O(\sqrt{n}).
\]
The expected space usage of the first phase is
$O(n^{3/2})$. Moreover, $S_i$ is independent of
$G_i$, so by Lemma~\ref{lem:4do-edge-bound},
\[
\mathbb{E}[|E(G_i)|\cdot |S_i|]
=
\mathbb{E}[|E(G_i)|]\cdot\mathbb{E}[|S_i|]
=O(n^2).
\]
Summing over all iterations, the expected time for
the Dijkstra computations is $\widetilde O(n^2)$.

In addition, in each iteration $i$ and for each vertex $v$, the algorithm computes
$\min_{s\in S_i} d_i(v \lra s)$ in order to select a pivot.
This requires $O(n|S_i|)$ time per iteration, and therefore contributes a total expected
time of $O(n^{3/2})$ over all iterations.
\end{proof}

\section{Using FMM to Accelerate the Algorithms}
\label{app:fmm}

\paragraph{Min-Plus Product}

We begin by formalizing the min-plus product.

\begin{definition}[Min-plus matrix product]
Let $A \in (\mathbb{R}_{\ge 0} \cup \{\infty\})^{n \times t}$ and
$B \in (\mathbb{R}_{\ge 0} \cup \{\infty\})^{t \times m}$. Their
\emph{min-plus product} $C = A \star B$ is the matrix
$C \in (\mathbb{R}_{\ge 0} \cup \{\infty\})^{n \times m}$ defined by
\[
 C[i,j]
 =
 \min_{1 \le \ell \le t}
 \bigl(A[i,\ell] + B[\ell,j]\bigr).
\]
\end{definition}

We denote by $\omega(1,\kappa,1)$ the algebraic exponent for multiplying an
$n\times n^\kappa$ matrix by an $n^\kappa\times n$ matrix in
$O(n^{\omega(1,\kappa,1)})$ time. We use the following standard black box:
combining Zwick's scaling-based $(1+\varepsilon)$-approximate distance-product
framework~\cite{Zwick2002} with bounds for rectangular matrix multiplication
(see, e.g.,~\cite{Alman25}), one can compute a one-sided
$(1+\varepsilon)$-approximate rectangular min-plus product of an
$n\times n^\kappa$ matrix and an $n^\kappa\times n$ matrix, whose finite
entries lie in $[1,W]$, in time
\[
 O\!\left(
 n^{\omega(1,\kappa,1)}
 \frac{\log(Wn)}{\varepsilon}
 \right).
\]
That is, if $\widetilde C$ is the computed approximation to $C=A\star B$,
then
\[
 C[i,j]
 \le
 \widetilde C[i,j]
 \le
 (1+\varepsilon)C[i,j]
\]
for every $i,j$. Zero and infinite entries can be handled separately in the
standard way.

\subsection{$(4+\varepsilon)$-APSP}
\label{subsec:4-algebraic-speedup}

The fully combinatorial algorithm of Section~\ref{sec:4-apsp} spends
$O(n^2|S_i|)$ time in iteration $i$ to compute
\[
 \min_{s\in S_i}
 \bigl\{
 d_i(v \lra s) + d_i(s \lra u)
 \bigr\}
\]
by explicit enumeration over $s\in S_i$. This is the bottleneck in the
aggregation step of the sparsification procedure.

To accelerate this aggregation, assume that $w(e)\in[1,W]$ for every
$e\in E$, and fix $\varepsilon\in(0,1]$. Let
\[
 A_i(v,s):=d_i(v\lra s),
\]
so that $A_i$ is an $n\times |S_i|$ matrix. Since roundtrip distance is
symmetric, the desired values are given by the min-plus product
\(
 A_i\star A_i^\top.
\)
We replace the explicit enumeration by a
$(1+\varepsilon/4)$-approximate rectangular min-plus product
\(
 B_i \approx A_i\star A_i^\top.
\)
The finite entries of $A_i$ lie in $[1,O(nW)]$, apart from zero diagonal
entries, which are handled separately. Thus, for every $v,u\in V$,
\[
\begin{aligned}
 \min_{s\in S_i}
 \bigl\{
 d_i(v\lra s)+d_i(s\lra u)
 \bigr\}
 &\le
 B_i(v,u) \\
 &\le
 \left(1+\frac{\varepsilon}{4}\right)
 \min_{s\in S_i}
 \bigl\{
 d_i(v\lra s)+d_i(s\lra u)
 \bigr\}.
\end{aligned}
\]
The remainder of the algorithm---sampling, Dijkstra computations, and edge
pruning---is unchanged.

Suppose that the shortest roundtrip between $v$ and $u$ is destroyed by the
sparsification. By Lemma~\ref{lem:roundtrip-stretch-general}, for some
iteration $i$,
\[
 \min_{s\in S_i}
 \bigl\{
 d_i(v\lra s)+d_i(s\lra u)
 \bigr\}
 \le
 4d_G(v\lra u).
\]
Consequently,
\[
 B_i(v,u)
 \le
 \left(1+\frac{\varepsilon}{4}\right)4d_G(v\lra u)
 =
 (4+\varepsilon)d_G(v\lra u).
\]
If the shortest roundtrip survives in the final graph $H$, the
$3$-approximation algorithm run on $H$ gives a valid
$(4+\varepsilon)$-approximation. Hence the resulting approximation guarantee
is $4+\varepsilon$.

Since the sizes of the sets $S_i$ grow geometrically and are at most
$\widetilde O(n^\beta)$, the total time for all approximate min-plus products
is
\[
 \widetilde O\!\left(
 n^{\omega(1,\beta,1)}
 \frac{\log(nW)}{\varepsilon}
 \right).
\]
All other work in the sparsification procedure takes
$\widetilde O(n^2)$ expected time. Moreover,
\[
 \mathbb{E}[|E(H)|]
 =
 \widetilde O(n^{2-\beta}),
\]
and therefore running the $3$-approximation algorithm on $H$ takes
\[
 \widetilde O\!\left(
 |E(H)|\sqrt n+n^2
 \right)
 =
 \widetilde O\!\left(
 n^{5/2-\beta}+n^2
 \right)
\]
expected time.

The total expected running time is therefore
\[
 \widetilde O\!\left(
 n^{\omega(1,\beta,1)}
 \frac{\log(nW)}{\varepsilon}
 +
 n^{5/2-\beta}
 +
 n^2
 \right).
\]
Choosing $\beta$ so that
\[
 \omega(1,\beta,1)
 =
 \frac{5}{2}-\beta
\]
and using the current bounds for rectangular matrix multiplication gives the
stated
\[
 \widetilde O\!\left(
 n^{2.032}\frac{\log(nW)}{\varepsilon}
 \right)
\]
running time.

\subsection{$(3+\varepsilon)$-APSP}
\label{subsec:3-algebraic-speedup}

As in Section~\ref{subsec:4-algebraic-speedup}, we accelerate the aggregation
through the sampled vertices. Here, we retain the exact incoming and outgoing
Dijkstra searches from the sampled vertices, and accelerate only the
all-pairs aggregation.

We use the algorithm of Section~\ref{sec:fast-3} with a single sampling
exponent $\alpha$. As in that section, centers and pivots are sampled
independently, with $\widetilde O(n^\alpha)$ sampled vertices of each type.
Let
\[
S:=\bigcup_i(S_i^c\cup S_i^p),
\qquad
|S|=\widetilde O(n^\alpha).
\]

For every $s\in S$, we run incoming and outgoing Dijkstra searches in
the original graph $G$, computing
\[
d(v\to s)
\qquad\text{and}\qquad
d(s\to v)
\]
for every $v\in V$. This takes $\widetilde O(mn^\alpha)$ time.
Since all distances used in the pivot and center comparisons are exact,
the size analysis of Section~\ref{sec:fast-3} remains unchanged.
Thus, for every $v\in V$,
\[
\mathbb{E}[|N_H(v)|],
\mathbb{E}[|T_H(v)|]
=
\widetilde O(n^{1-\alpha}).
\]

The all-pairs proxy values through the sampled centers are computed
using a $(1+\varepsilon/3)$-approximate rectangular min-plus product,
as in Section~\ref{subsec:4-algebraic-speedup}.
For the pivots, instead of aggregating through all sampled vertices,
we update, for every $v,u\in V$,
\[
\delta(v,u)\gets
\min\left\{
\delta(v,u),
\min_{p\in P(v)}
\{d(v\to p)+d(p\to u)\}
\right\}.
\]
Since $|P(v)|=O(\log n)$, these updates take $O(n^2\log n)$ time.
The final truncated searches are unchanged.

We now analyze the approximation error. As in Lemma~5.1, consider
a shortest path from $v$ to $u$. If the path is destroyed by
sparsification, some sampled center $s$ satisfies
\[
d(v\to s)+d(s\to u)
\le 2d(v\to u)+d(u\to v).
\]
The approximate min-plus product therefore gives
\[
\delta(v,u)
\le
\left(1+\frac{\varepsilon}{3}\right)
\bigl(2d(v\to u)+d(u\to v)\bigr).
\]
If the path survives but the truncated search stops, a pivot in
$P(v)$ provides the same bound without the additional approximation
factor. Otherwise, the exact distance is found.
Consequently,
\[
\begin{aligned}
\delta(v,u)+\delta(u,v)
&\le
\left(1+\frac{\varepsilon}{3}\right)
3d(v\lra u)\\
&=(3+\varepsilon)d(v\lra u).
\end{aligned}
\]

The Dijkstra searches take $\widetilde O(mn^\alpha)$ time, while
the approximate min-plus product takes
\[
\widetilde O\!\left(
n^{\omega(1,\alpha,1)}
\frac{\log(nW)}{\varepsilon}
\right)
\]
time. The final truncated searches take
\[
\widetilde O\!\left(
n\cdot n^{1-\alpha}\cdot n^{1-\alpha}
\right)
=
\widetilde O(n^{3-2\alpha}).
\]
Thus, the total expected running time is
\[
\widetilde O\!\left(
mn^\alpha
+
n^{\omega(1,\alpha,1)}
\frac{\log(nW)}{\varepsilon}
+
n^{3-2\alpha}
+
n^2
\right).
\]

Let $\alpha_0$ satisfy
\[
\omega(1,\alpha_0,1)=3-2\alpha_0\le 2.037,
\]
using the current bounds for rectangular matrix multiplication
(see, e.g.,~\cite{Alman25}).
We choose
\[
n^\alpha=\min\left\{n^{\alpha_0},
\frac{n}{m^{1/3}}\right\}.
\]
If $m\le n^{3(1-\alpha_0)}$, then $\alpha=\alpha_0$,
and all three terms are bounded by
$\widetilde O(n^{2.037}\log(nW)/\varepsilon)$.
Otherwise, $n^\alpha=n/m^{1/3}$, and the Dijkstra searches
and truncated searches both take $\widetilde O(nm^{2/3})$ time.
Since $\alpha\le\alpha_0$, the matrix multiplication term
is dominated by this bound.

We therefore obtain a $(3+\varepsilon)$-approximation in expected time
\[
\widetilde O\!\left(
(n^{2.037}+nm^{2/3})
\frac{\log(nW)}{\varepsilon}
\right).
\]
For $m\le n^{3/2}$, Corollary~\ref{cor:3-apsp}
already gives a stretch-$3$ approximation in
$\widetilde O(n^2)$ time.

\end{document}